\documentclass[aps, prd, a4paper, 10pt, twocolumn, nofootinbib, showkeys]{revtex4-2}

\usepackage[margin=2.5cm]{geometry}
\usepackage{graphicx}
\usepackage{xcolor}
\usepackage{braket}
\usepackage{multirow}
\usepackage{amsmath, amsfonts, dsfont}
\usepackage{enumitem}
\usepackage{tikz}
\usetikzlibrary{matrix, arrows.meta}

\renewcommand{\j}{I}
\newcommand{\J}{T}
\newcommand{\I}{I}

\begin{document}

\title{Symmetries in particle physics:\\ from nuclear isospin to the quark model}
\author{Bruno Berganholi}
\author{Gl\'auber C. \surname{Dorsch}}
\email{glauber@fisica.ufmg.br}
\author{Beatriz M. D. \surname{Sena}}
\author{Giovanna F. do Valle}
\affiliation{\vskip2mm Departamento de F\'isica\\ Universidade Federal de Minas Gerais (UFMG)\\ Belo Horizonte, MG, Brazil}
\begin{abstract}
    We present a concise pedagogic introduction to group representation theory motivated by the historical developments surrounding the advent of the Eightfold Way. Abstract definitions of groups and representations are avoided in favour of the physical intuition of symmetries of the nuclear interaction. The concept of nuclear isospin is used as a physical motivation to introduce SU(2) and discuss the main techniques of representation theory.
    The discovery of strange particles motivates extending the symmetry group to SU(3), at first in the context of the Sakata model. We highlight the successes in fitting mesons in the SU(3) octet, discuss the drawbacks of the Sakata model for baryonic classifications, and how the Eightfold Way finally led to the quark model. This approach has two major advantages: (i) the main concepts of the theory of Lie groups are introduced and discussed without ever losing touch with its applications in particle physics; (ii) it allows the beginner to study group theory while also becoming acquainted with the historical developments of particle physics that led to the concept of quarks. In particular, in this pedagogical path the quarks appear as yet another class of particles predicted from symmetry principles, rather than being introduced \emph{ad hoc} for postulating an SU(3) symmetry, as usually done in the literature.
\end{abstract}
\maketitle 

\section{Introduction}

Throughout the history of particle physics, symmetry principles have proven to be extremely productive tools for model building. The recognition of an underlying symmetry of Nature allows one to predict properties of fundamental particles, to obtain quantitative predictions on relations between observables, and even to anticipate the precise forms of interaction terms in the Lagrangian. In fact, the entire edifice of the Standard Model and its attempted low-energy extensions rests upon symmetry principles such as \emph{gauge invariance}, a statement about local symmetries of the fundamental interactions. 
It is therefore essential that any novice in particle physics be fluent in the mathematical description of symmetries and their physical consequences. This is precisely the scope of the discipline called \emph{group representation theory}.

Symmetries of physical laws are most easily illustrated in the context of Lorentz covariance and relativity. In this case the transformation between inertial frames can be readily visualized, and representation theory can be motivated as a tool for building objects that transform in a well-defined way under a change of frame, allowing different observers to compare their experimental outputs. But the Lorentz group is certainly not the simplest, and the study of its representations relies on some previous knowledge of SU(2). On the other hand, more basic group structures could be introduced by appealing to internal symmetries, but attempting to use the Standard Model gauge groups as motivation makes the situation significantly murkier. To explain concepts such as weak isospin and the SU(2)$_L$ symmetry of weak interactions, one must invoke a notion of ``similarity'' between electrons and neutrinos, between up- and down-type quarks. But any such similarity is blatantly badly broken by the sharp discrepancy in these particles' masses. And it's not helpful to explain this as a manifestation of symmetry breaking by the Higgs mechanism: the elegance of this solution is only appreciated if one already grasped the importance of enforcing these symmetries in the first place. 
One can safely say that attempting to build an intuition of internal symmetries by invoking a case where it is badly broken is not a very pedagogic approach. The situation is improved a little if one starts from the SU$(3)_c$ case of chromodynamics, since this one is indeed preserved. However, this also requires introducing new concepts such as color charge, and has the drawback of using the more complicated SU$(3)$ group as a starting point.

An alternative and more pedagogic path is to discuss internal symmetries starting from their historical origins, namely Heisenberg's notion of nuclear isospin. For that, the only prerequisite are the concepts of protons and neutrons familiar to any undergraduate student. One can then use representation theory to predict the properties of the pions, and predict the ratio of nuclear reaction cross sections from Clebsch-Gordan coefficients. Then, by extending the symmetry group from SU$(2)$ to SU$(3)$, one can understand hadronic classification schemes, discuss the successful prediction of the $\Omega^-$ particle, and introduce the concept of quarks in their appropriate historical background.

While this approach can be found in many excellent textbooks on the subject, each has its own goals and emphasis, leaving out a good deal of important material. Some discuss much of the physics without going into detail on group theoretical developments, e.g. refs.~\cite{Halzen:1986, Griffiths:2008}. Others emphasize only the necessary group theory for the subsequent application to the Standard Model, leaving out important applications on nuclear physics, e.g.~\cite{Aitchison:2004cs}. Others, while providing a very thorough discussion of the history behind these symmetry principles, were written before the full development and confirmation of the quark model and hence would not discuss it fully~\cite{Gasiorowicz:1966xra}\footnote{It is also worth noticing that ref.~\cite{Gasiorowicz:1966xra} uses this hadronic classification approach to introduce the use of symmetries in particle physics not by pure choice, but also because the more fundamental internal symmetries of the weak and strong interactions were still unknown at the time of its writing.}.

There is also a plethora of pedagogic materials aimed at introducing group theory for particle physicists~\cite{Georgi:1999wka, Zee:2016fuk, Thyssen:2017}. Many of these also discuss the use of internal symmetries at the dawn of nuclear physics, but the physics is inevitably swamped in the dense formalism of generic group theoretical results. 

Our purpose in this paper is to offer a concise yet self-contained introduction to the main elements of group representation theory through its application to the hadronic classification schemes, starting with Heisenberg's idea of isospin and culminating in the Eightfold Way and the quark model.

Although the content presented here is not new, the novelty is in the presentation sequence. We have aimed for a healthy balance between the formal aspects of representation theory and the underlying developments in hadronic physics. In doing so we omitted formal definitions of group theory (some of which we put in the appendix), but without sacrificing the overall understanding of the main concepts and techniques. In fact, these formal definitions can even obscure rather than enlighten the topic to the very beginner. Though this presentation does not exhaust the topic, it should allow the reader to transition smoothly to other standard texts on the subject, including (but not restricted to) those cited above.

The paper is organized as follows. In section~\ref{sec:isospin} we motivate the concept of nuclear isospin, show that it is associated to the SU$(2)$ group, and build its representations. Then in section~\ref{sec:sum} we discuss rules for summing isospin, i.e. building representations related to composite states. This culminates in the Clebsch-Gordan coefficients, which we use to predict the ratio of certain nucleon + nucleon and pion + nucleon scattering cross-sections. In section~\ref{sec:su3} we extend the symmetry group to SU$(3)$ to account for the ``particle zoo'' of the 1940s and 50s and discuss its representations, culminating in the Eightfold Way and the quark model. Our conclusions are reserved for section~\ref{sec:conclusions}.

%%%%%%%%%%%%%%%%%%%%%%%%%%%%%%

\section{Nuclear isospin}
\label{sec:isospin}
\begin{figure}
    \centering
    \includegraphics[scale=.22]{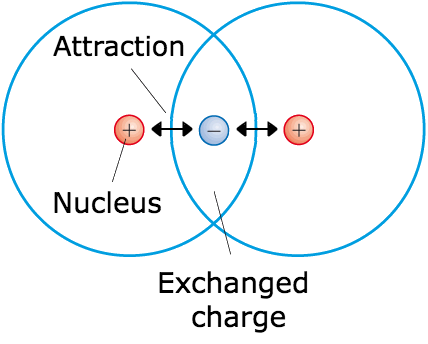}
    \qquad\quad
    \includegraphics[scale=.3]{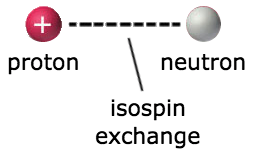}
    \caption{(Left) The covalent bond among the two protons in the $H_2^+$ cation is mediated by the exchange of an electron. This can only happen when the two electronic orbitals around the protons overlap, so the electron can be found in a superposition of both states. As the two protons are brought apart, the superposition becomes untenable and the interaction ceases: the associated force is short-ranged. (Right) Heisenberg intuited that the short-ranged nuclear force was of a similar nature, mediated by the exchange of a new type of charge.}
    \label{fig:H2}
\end{figure}

The discovery of the neutron in 1932, and the recognition of its role as a constituent of the atomic nucleus, immediately required a postulated new interaction responsible for keeping neutrons and protons bound together. The first concrete attempt at a description of this force was put forward by Heisenberg in his seminal 1932 series of papers~\cite{Heisenberg:1932dw, Heisenberg:1932II, Heisenberg:1932III}, merely a few months after the reports on the discovery of the neutron\footnote{See also~\cite{Kemmer:1982bj, Cassen:1936dg, Borrelli:201781} for examples and discussions on the early history of the concept of isospin in nuclear physics.}. Motivated by the short-ranged nature of the nuclear force, Heisenberg thought of this interaction as analogous to the covalent bond in the dihydrogen cation $H_2^+$~\cite{FeynmanLectures}, depicted in figure~\ref{fig:H2} (left). In this case, two protons are held together by the exchange of an electron, which takes place when the two orbitals overlap. As soon as the protons are separated by a distance larger than the orbital size, i.e. the Bohr radius $\sim 10^{-10}$~m, the exchange ceases and the electron becomes attached to one of protons only, turning it into a neutral $H$ atom which no longer interacts electromagnetically with the remaining $H^+$ ion. Heisenberg intuited that a similar mechanism was responsible for the short-ranged nuclear force, but involving the exchange of another type of charge altogether. 

Specifically, he postulated that the particles in the nucleus --- so-called \emph{nucleons} --- carry another charge whose value characterizes them as a proton or a neutron. The nucleon is thus a two-level system, analogous to a spin-1/2 particle: the $+1/2$ state corresponds to a proton, the $-1/2$ to a neutron. Due to this similarity with spin, this charge is called \emph{isospin}\footnote{In the beginning people called it isotopic spin. But in reality the transformation protons $\leftrightarrow$ neutrons relates isobar nuclei, rather than isotopes, so other people called it isobaric spin. Eventually the notation was simplified to isospin.}. 

\subsection{Charge independence of the nuclear force}

Now, protons and neutrons have remarkably near degenerate masses~\cite{ParticleDataGroup:2022pth}, 
\begin{equation*}\begin{split}
    m_p &= 938.27208816 \pm 0.00000029~\text{MeV},\\
    m_n &= 939.5654205 \pm 0.0000005~\text{MeV},
\end{split}\end{equation*}
amounting to a relative mass difference
\[ \frac{m_n-m_p}{\left(\frac{m_n+m_p}{2}\right)} \approx 0.14\%.\]
This means that a ``flip'' in the isospin of a free nucleon involves a comparably negligible amount of energy, so that free protons and free neutrons are approximately degenerate states. 

\begin{figure}
    \centering
    \includegraphics[scale=1.1]{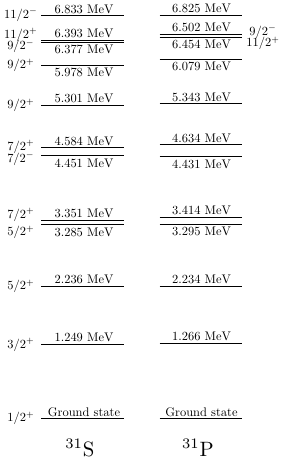}
    \caption{Energy levels of two mirror nuclei $^{31}$S and $^{31}$P. The energy of each level relative to the ground state is shown, together with its spin and parity. These two nuclei are related to each other by an exchange of protons $\leftrightarrow$ neutrons. Slight differences in the energy levels of these nuclei are well explained by electromagnetic effects~\cite{Jenkins:2005jk}. The fact that these levels nearly coincide for these two nuclei indicates that exchanging protons and neutrons has little impact on the nuclear force.}
    \label{fig:levels}
\end{figure}

As for when they are interacting inside the nucleus, as soon as more information was gathered on the phenomenology of this nuclear force, it gradually became clear that this interaction did not seem to distinguish protons from neutrons. This is known as the approximate ``charge independence'' of the nuclear force. One evidence for this comes from the similar properties of proton-proton, proton-neutron and neutron-neutron scatterings (discounting for electromagnetic effects, which affect each of these processes differently)~\cite{Machleidt:2001rw, Miller:2006tv}. Quantitatively, the nuclear potentials describing these three processes differ from one another by no more than $\sim 2.5\%$~\cite{Miller:2006tv}. Another evidence comes from the strikingly similar energy levels of mirror nuclei, such as $^{31}$P and $^{31}$S, as shown in figure~\ref{fig:levels}. These nuclei are related to one another by a proton $\leftrightarrow$ neutron exchange\footnote{The $^{31}$P nucleus has 15 protons and 16 neutrons, whereas $^{31}$S has 16 protons and 15 neutrons.}. Small differences in energy levels of these nuclei are well accounted for by electromagnetic effects~\cite{Jenkins:2005jk}, which again shows that the nuclear interaction (which is dominant inside the nucleus) does not sharply differentiate protons and neutrons. 

Altogether, this means that \emph{a ``rotation'' in isospin space is an approximate symmetry of nature}. By this we mean that there exist certain transformations that can be performed over the physical system while keeping its dynamics completely unaltered. The system before and after the transformation are completely identical. The mere recognition of this fact has deep physical implications, as we will see in the following.

\subsection{Isospin and SU(2)}
\label{sec:isospinsu2}

Let us put the above discussion in more formal terms. We have argued above that the nuclear interaction does not seem to distinguish protons from neutrons\footnote{Obviously the electromagnetic interaction is able to distinguish them, but its effects are subdominant compared to the nuclear force, which is typically much stronger.}. From the previous discussion, we can postulate that, as far as the nuclear interaction is concerned, the proton and the neutron behave as the same kind of particle, which we call the \emph{nucleon} (i.e. a particle of the atomic nucleus). This nucleon can be found in two different states in Nature,
\begin{equation*}
    p = \begin{pmatrix} 1\\0 \end{pmatrix},
    \qquad
    n = \begin{pmatrix} 0\\1 \end{pmatrix},
\end{equation*}
corresponding to ``isospin up'' and ``isospin down'' states, respectively. 

In general, a nucleon state can be found as a superposition described by
\begin{equation*}
    N = \begin{pmatrix} \psi_p\\ \psi_n \end{pmatrix} = \psi_p \,p + \psi_n\, n.
\end{equation*}
Isospin symmetry means that any redefinition of $p$ and $n$ describes an equivalent physics. 
This transformation can be described via the matrix
\begin{equation}
   N^\prime 
   = U N, \quad\text{with}\quad U = \begin{pmatrix} \alpha & \beta \\ \gamma & \delta\end{pmatrix} .
    \label{eq:U}
\end{equation}
Since the physics is unaltered by this transformation, all probabilities must be the same  when calculated with primed or unprimed states. In particular, 
\begin{equation*}
    N^{\dagger} N = N^{\prime\,\dagger} N^\prime = N^\dagger U^\dagger U N.
\end{equation*}
This implies that $U$ must be unitary,
\begin{equation*}
    U^\dagger U = \mathds{1}_{2\times 2}.
\end{equation*}

Consider now the matrix determinant
\begin{equation*}\begin{split}
    1 &= \det \mathds{1}_{2\times 2} = \det (U^\dagger U) \\
       & = (\det U^\dagger) \det U 
       = |\det U|^2 .
\end{split}\end{equation*}
Since $U$ is a complex matrix, it follows that the determinant of $U$ is an overall complex phase, $\det U = \exp(i\theta)$
for some $\theta \in \mathbb{R}$. It is well known that multiplying a quantum state by an overall complex phase has no physical implications, so we can restrict outselves to transformations with $\det U=1$. Put another way, the only novelty in eq.~\eqref{eq:U} lies in the invariance of physics under a $2\times 2$ unitary transformation with $\det U=1$. The set of matrices satisfying these conditions forms a group called SU(2)\footnote{This is an example of a ``special unitary'' group. The ``S'' stands for ``special'', meaning unitary determinant. The ``U'' stands for ``unitary matrices'' and the ``2'' means the matrices are $2\times 2$.}.

The transformation described in eq.~\eqref{eq:U} is analogous to a ``rotation'' in isospin space. In other words, it is a redefinition of what we call ``up'' and ``down'' or ``proton'' and ``neutron'' states (in much the same way that a spacetime rotation is a redefinition of what we call $x, y$ and $z$ axes). This is an example of an \emph{internal symmetry} of nature, so-called because the transformation does not act on spacetime coordinates (like usual spatial rotations), but involves field redefinitions. The symmetry statement, in this case, is that the nuclear interaction does not distinguish protons and neutrons, so a redefinition of these fields (a re-orientation of directions in this field space) has no impact on the physical results.

The recognition that a ``rotation'' in isospin space is a symmetry of the nuclear interaction has profound consequences. For example, consider a particle that carries some isospin, and therefore interacts via the nuclear force. If we perform an isospin redefinition, the nature of this particle, as far as the nuclear force is concerned, must remain unchanged. The interaction treats it as the same kind of particle, only in a different isospin state (otherwise this transformation wouldn't be a symmetry). Consequently, each type of particle belongs to a vector space where the state vectors can be related to one another by some isospin transformation. In other words, the transformation does not map a state to another belonging to a different vector space, which would represent a different kind of particle.

This statement is so important that it is worth phrasing it another way. If we redefine isospin, there must be a dictionary that allows us to find the ``new'' state of this particle in terms of the pre-transformed one. Only then would we be able to relate the physics under this new definition to the previous description and make sure that it remains indeed unaltered\footnote{Thinking about spacetime symmetries could make this argument simpler to grasp. Suppose $\mathcal{S}$ is an inertial observer at rest with respect to some electric charge, and $\mathcal{S}^\prime$ sees it moving with constant velocity. Clearly $\mathcal{S}$ sees only a radial electric field emanating from the charge, while $\mathcal{S}^\prime$ sees also a current and therefore a magnetic field as well. They disagree as to the intensity of the electric and magnetic fields, as well as to the charge and current distribution. But there must be a way to translate one's observations into the other's, for only then can they mutually communicate about their experiments and reach an agreement as to the universality of Maxwell's equations.}. The state space of a particle is the space of all objects that can be transformed into one another by an isospin transformation. Technically, this means that physical states must belong to \emph{representations} of the SU(2) symmetry group.

We have already seen an instance of this, when we said that ``proton'' and ``neutron'' are two different states of the nucleon, which means that the space of nucleon states is two-dimensional. Other particles may belong to higher-dimensional representations.

How do we construct these representation spaces? Starting from one reference state $|\psi\rangle$, we can apply an SU(2) transformation (an isospin redefinition), mapping this state onto some $|\psi^\prime\rangle$. This new state still describes the same particle, only in a different isospin configuration. 
Then perform another isospin redefinition, taking us from $|\psi^\prime\rangle$ to $|\psi^{\prime\prime}\rangle$, and so on. If the state space of such particle is finite dimensional, eventually this procedure will cease to produce linearly independent vectors. Thus, by performing all possible symmetry transformations on a reference state, one can obtain the whole state space of the system. By construction all transformations will map this space onto itself, and all states are relatable to one another by one such transformation: then we will have built a representation.

The remaining question is: what does an isospin redefinition looks like when acting on higher-dimensional spaces? Eq.~\eqref{eq:U} reveals only how it affects the two-dimen\-sional (isospin $1/2$) representation. Before trying to generalize it, it is worth looking into this representation in more detail.

\subsection{Infinitesimal transformations and the two-dimensional representation}
\label{sec:su2inftl}

The group SU(2) has a structure that simplifies enormously our task of finding representations: it is a group of \emph{continuous} transformations, meaning that we can ``rotate'' in isospin space by an infinitesimal amount. Think about infinitesimal parameters in equation~\eqref{eq:U}, for instance. Such a continuous group is called a \emph{Lie group}. Because of this, instead of having to perform all possible rotations on our reference state $|\psi\rangle$ to build the representation space, we need only perform infinitesimal rotations along independent ``directions''. 

What does the matrix $U$ in eq.~\eqref{eq:U} looks like for infinitesimal transformations? By definition this is very close to ``doing nothing'' (we are rotating by tiny, insignificant ``angles''), so the transformation matrix must be close to the identity. Indeed, we could perform a Taylor expansion and, neglecting terms of quadratic order or higher, one expects
\begin{equation}
   U = \mathds{1}_{2\times 2} + i\omega,
   \qquad \omega = \begin{pmatrix}
        \omega_{11} & \omega_{12}\\ \omega_{21} & \omega_{22}
    \end{pmatrix}.
\end{equation}
Since $\det U = 1$, 
\begin{equation*}\begin{split}
    1 &= \det (\mathds{1}_{2\times 2} + i\omega) \\
    &= 1 + i(\omega_{11} + \omega_{22}) - (\omega_{11}\omega_{22} + \omega_{12}\omega_{21}). 
\end{split}\end{equation*}
Neglecting terms $\mathcal{O}(\omega^2)$ (because $\omega$ is infinitesimal) one finds
$
    \omega_{11} + \omega_{22} \equiv \text{tr}(\omega) = 0.
$
Moreover, since $U$ is unitary,
\begin{equation*}\begin{split}
    U^\dagger U = \mathds{1}_{2\times 2} & = (\mathds{1}_{2\times 2} +  i\omega)(\mathds{1}_{2\times 2} - i\omega^\dagger)\\
    &= \mathds{1}_{2\times 2} + i (\omega - \omega^\dagger) + \mathcal{O}(\omega^2),
\end{split}\end{equation*}
hence\footnote{We have chosen to define $U=\mathds{1}+i\omega$, with this $i$ factor in front, precisely to enforce that $\omega$ be hermitean. Without this factor it would be anti-hermitean, which wouldn't change much of the conclusions below, but would introduce some inconvenient factors of (-1).}
\begin{equation}
    \omega = \omega^\dagger.
    \label{eq:Jhermitean}
\end{equation}

Thus $\omega$ is a hermitean traceless matrix, and can be put in the form
\begin{equation*}
    \omega = \frac{1}{2}\begin{pmatrix}
        d\theta^3 & d\theta^1-id\theta^2 \\ d\theta^1+id\theta^2 & -d\theta^3
    \end{pmatrix}
    = {d\vec{\theta}\cdot \vec{\J}},
\end{equation*}
with $\vec{\J}=(\J_1, \J_2, \J_3)$ are (half) the Pauli matrices,
\begin{equation}\begin{split}
    \J_1 &=\frac12
    \begin{pmatrix}
        0&1\\1&0
    \end{pmatrix}, \quad 
    \J_2 =
    \frac12\begin{pmatrix}
        0&-i\\ i&0
    \end{pmatrix},\\
     & \qquad\quad 
     \J_3 =
    \frac12\begin{pmatrix}
        1&0\\ 0&-1
    \end{pmatrix}.
    \label{eq:Js2d}
\end{split}\end{equation}

We see that an isospin transformation is parameterized by 3 angles, just like a rotation in 3-dimensional space\footnote{This is an aspect of the famous results that SU(2) is homomorphic to (i.e. has the same group structure as) SO(3), the group of orthogonal $3\times 3$ matrices that describe rotations in 3 dimensions.}. The $\J_1, \J_2$ and $\J_3$ are called generators of infinitesimal rotations, and correspond to transformations along these three independent directions.

Notice that $\J_3$ is an operator whose eigenvectors are the $p$ and $n$ states. It tells us the isospin configuration of the nucleon in a certain state. On the other hand $\J_1$ and $\J_2$ can be combined into
\begin{equation}
    \J_1+i\J_2 = \begin{pmatrix} 0 & 1\\ 0 & 0\end{pmatrix}
    \quad\text{and}\quad
    \J_1-i\J_2 = \begin{pmatrix} 0 & 0\\ 1 & 0\end{pmatrix},
\end{equation}
which act on the nucleon states such that
\begin{equation}\begin{split}
    (\J_1+i\J_2)\, n &= p,\\
    (\J_1 - i\J_2)\, p &=n,
\end{split}\end{equation}
i.e. they respectively raise and lower the value of isospin by one unit. For this reason they are called ``raising'' and ``lowering'' operators.

Thus the state space of nucleons can be constructed by starting from the $p$ state and continuously applying the lowering operator. Applying it once yields $n$. Applying it again on $n$ yields the zero vector: no new state is obtained and the procedure stops. Thus the space of nucleon states is two-dimensional.

\subsection{The algebra of SU(2) and other finite dimensional representations}
\label{sec:repssu2}

The above procedure can be generalized for higher-dimensional spaces in the following way. Regardless of their specific shape in $n$ dimensions, the three generators of SU(2) are characterized by the commutation relations they satisfy\footnote{The reader can easily show that the matrices in eq.~\eqref{eq:Js2d} do satisfy these relations.}, namely
\begin{equation}
    [\J_i, \J_j] = i\epsilon_{ijk} \J_k.
    \label{eq:su2algebra}
\end{equation}
Any three matrices that satisfy these relations are intimately related to SU(2) transformations, and are said to belong to the $\mathfrak{su}(2)$ algebra\footnote{The algebra is the vector space of generators of infinitesimal transformations, together with a commutation relation. The group structure is characterized by such commutation relation of the generators. SU(2) is characterized by relations~\eqref{eq:su2algebra}, with $\epsilon_{ijk}$ as coefficients of the algebra, called \emph{structure constants}. Other commutation relations give rise to different groups, as we will see below.}.

Since these generators do not commute, it is impossible to diagonalize all three of them simultaneously. But one can show from~\eqref{eq:su2algebra} that
\begin{equation*}
    \J^2\equiv \J_1^2 + \J_2^2 + \J_3^2
\end{equation*}
commutes with all three $\J$'s. So we choose to work with normalized states $|\j,m\rangle$ that diagonalize $\J_3$ and $\J^2$ simultaneously, 
\begin{equation}\begin{split}
    \J_3 |\j,m\rangle &= m |\j,m\rangle,\\
    \J^2 |\j,m\rangle &= \j(\j+1)|\j,m\rangle.
\end{split}\end{equation}
It will soon become clear why we chose to write the eigenvalues of $\J^2$ in this funny way.

Instead of working with $\J_1$ and $\J_2$ we will repeat the argument of the previous section and define
\begin{equation*}
    \J_{\pm} = \J_1 \pm i \J_2.
\end{equation*}
From eq.~\eqref{eq:su2algebra} it is easy to show that
\begin{equation}\begin{split}
    [\J_3,\J_{\pm}]=\pm \J_{\pm},
    \quad\text{and}\quad
    [\J_{+},\J_{-}]=2\J_3.
    \label{eq:J3pm_algebra}
\end{split}\end{equation}
This then means that
\begin{equation}\begin{split}
    \J_3 (\J_\pm |\j,m\rangle) &= ([\J_3, \J_\pm] + \J_\pm \J_3)|\j,m\rangle\\
    &= (m\pm1)(\J_\pm |\j,m\rangle).
    \label{eq:mpm1}
\end{split}\end{equation}
In other words, $\J_\pm$ raises/lowers the eigenvalue of $\J_3$ by one unit. But $\J_\pm$ are related to rotations in the $\J_1$ and $\J_2$ directions, so the above procedure teaches us how to obtain other vectors by rotating $|\j,m\rangle$. According to our argument in the previous section, all these vectors are states of the same particle. Notice that performing these rotations/isospin transformations alters $m$ but leaves $\j$ invariant. Therefore $\j$ labels the representation we are in (i.e. it labels the particle type), whereas $m$ labels a state of this particle. This makes sense because $\j$ is related to the total isospin $\J^2$, whereas rotations affect only the direction (thus altering only the components, say the $\J_3$ eigenvalue) but leave the vector length unchanged.

The idea is to find all states of the $\j$ representation by starting from the state with highest $m$ and applying $\J_-$ many times in succession. Each such application will give a different, linearly independent state\footnote{Recall from eq.~\eqref{eq:Jhermitean} that the $\J_i$'s are hermitean. Thus eigenvectors corresponding to different eigenvalues are orthogonal.}. But since we are in a finite dimensional space, this process must eventually end, i.e. there is a state $|\j,m_\text{min}\rangle$ such that $\J_-|\j,m_\text{min}\rangle=0$. 

We can find the value of $m_\text{min}$ (and therefore the dimension of the representation) in the following manner. From~\eqref{eq:mpm1} we know that
\begin{equation*}
    \J_\pm|\j,m\rangle = c^{\j,m}_\pm |\j,m\pm 1\rangle.
\end{equation*}
The proportionality constant $c_\pm^{\j,m}$ can be found by taking the norm of both sides and noting that $\J_+^\dagger = \J_-$,
\begin{equation*}
    \langle \j,m| \J_\mp \J_\pm |\j,m\rangle = |c_\pm^{\j,m}|^2.
\end{equation*}
But from the definition of $\J_\pm$ it follows that
$
    \J_\mp \J_\pm 
    = \J_1^2 + \J_2^2 \mp \J_3
    = \J^2 - \J_3 (\J_3 \pm 1),
$
so except for a phase that can be absorbed into the definition of $|\j,m\rangle$ one has
\begin{equation}
    \J_\pm\ket{j,m} = \sqrt{\j(\j+1) - m(m\pm 1)} \, \ket{\j,m\pm1}.
    \label{eq:cpm}
\end{equation}

Now, since we cannot raise $m$ to larger values than its maximum, we must have $\J_+|\j,m_\text{max}\rangle=0 \implies c_+^{\j,m_\text{max}}=0 \implies m_\text{max}=\j$~\footnote{That $\j=m_\text{max}$ follows from the choice of writing eigenvalues of $\J^2$ as $\j(\j+1)$, which thus justifies this (at first sight weird) choice.}. Likewise, we cannot decrease $m$ below the minimum $m_\text{min}$, and this then implies $m_\text{min}=-\j$. Therefore $m_\text{max}-m_\text{min} = 2\j$. But this difference must be a natural number (because we obtain $m_\text{min}$ by applying a number of $\J_-$'s to $|\j,m_\text{max}\rangle$). Thus the representations of SU(2) are characterized by half-integers
\begin{equation}
    \j = 0, \frac12, 1, \frac32, 2,\ldots,
    \label{eq:js}
\end{equation}
and each has dimension $2\j+1$ corresponding to states
\begin{equation}
    m = -\j, -\j+1,\ldots, \j-1, \j.
\end{equation}

It is now worth pausing to summarize these achievements. First, note that all of the above developments follow from the commutation relations of the generators of infinitesimal transformations, eq.~\eqref{eq:su2algebra}. This is a general result that underlies the theory of Lie groups: we can build representations just be looking at how states transform under infinitesimal transformations. In other words, we can focus almost exclusively on the Lie algebra.

Second, we see that the existence of a symmetry allows us to partition the entire state space of our theory into a number of independent subspaces. The procedure of finding representations of the symmetry group can then be seen as a way of organizing the total Hilbert space of states of a system. Indeed, we have defined the concept of ``representation space'' such that, under a symmetry transformation, the states of each of these subspaces transform only among themselves: there is no isospin rotation that can take a state in the $\j$ space to a different $\j^\prime$. Thus the transformation matrix acting on the total state space can be decomposed in block diagonal form\footnote{The $U_{\j=1/2}$ submatrix is the one given in eq.~\eqref{eq:U}.},
\begin{equation}
    U = 
    \left(\begin{array}{ccccccc}
        \multicolumn{1}{c|}{U^{\j=0}} & & & & & &\\ \cline{1-3}
	& \multicolumn{2}{|c|}{\multirow{3}{*}{$U^{\j=1/2}$}} & & & & \\ 
        & \multicolumn{2}{|c|}{}  & & & & \\
        & \multicolumn{2}{|c|}{}  & & & & \\ \cline{2-6}
        & & & \multicolumn{3}{|c|}{} &\\
        & & & \multicolumn{3}{|c|}{} & \\
        & & & \multicolumn{3}{|c|}{\raisebox{5pt}{~\quad $U^{\j=1}$\quad~~}} & \\
        & & & \multicolumn{3}{|c|}{} & \\  \cline{4-7}
        & & & & & & \multicolumn{1}{|c}{\ddots} \\ 
	\end{array}\right).
    \label{eq:bigU}
\end{equation}
The entire Hilbert space is said to be reducible. The subspaces $\j=0,1/2, 1,\ldots$, on the other hand, are \emph{irreducible} representations, because (by construction) they cannot be reduced any further (they have been built in such a way that every state is relatable to another by a transformation). Each of these irreducible subspaces correspond to different particle types as seen by the nuclear interaction.

\subsection{Symmetries and conservation laws}
\label{sec:conservation}

There is more. The very definition of isospin transformations being a symmetry of nature implies that such transformations do not affect the energy of the system. This means that, if $H$ is the hamiltonian of the system, with eigenstates $\ket{\psi_n}$, then $U\ket{\psi_n}$ is still an eigenstate with same eigenvalue, so
\begin{equation*}
    H (U\ket{\psi_n}) = E_{n} U\ket{\psi_n} 
    = U H\ket{\psi_n}. 
\end{equation*}
Since the $\ket{\psi_n}$ form a basis of the state space (because $H$ is hermitian) is follows that $[H,U]=0$, i.e. the hamiltonian commutes with the transformation operations. This means we can diagonalize both operators simultaneously, i.e. the $\ket{\j,m}$ states are also eigenstates of $H$, with definite energy. Therefore, the existence of a symmetry facilitates our task of finding solutions to the Schr\"odinger equation, since we may restrict ourselves to these subspaces to begin with\footnote{This is what we do when solving the hydrogen atom (or any problem with spherical symmetry) with separation of variables: the angular equation will always result in the spherical harmonics, which belong to (infinite dimensional) representations of the rotation group, and the problem reduces to solving the radial equation only.}.

Moreover, if $\j\neq \j^\prime$ then 
\[\langle \j^\prime, m^\prime| H|\j,m\rangle \sim \langle \j^\prime, m^\prime|\j,m\rangle = 0,\]
since states belonging to different representations are eigenstates of a hermitean operator $\J^2$ with different eigenvalues and are therefore orthogonal. But the left-hand side of this equation is the transition amplitude between states with $\j$ and $\j^\prime$. In other words, it is the probability amplitude that a system starts in a state in the $\j$ representation and is later\footnote{The Schr\"odinger equation can be written as $H|\psi(t)\rangle = i\frac{\partial}{\partial t}|\psi(t)\rangle \simeq i\frac{|\psi(t+\delta t)\rangle - |\psi(t)\rangle}{\delta t}$, or $|\psi(t+\delta t)\rangle = (\mathds{1}-i\delta t H)|\psi(t)\rangle$. We then see that $H$ is the generator of infinitesimal time translations, in much the same way as the $J_i$ are generators of infinitesimal isospin rotations, as discussed above.} found with total isospin $\j^\prime$. It follows that, if isospin is a symmetry of the hamiltonian, then \emph{the total isospin is conserved}\footnote{This is a particular example of a general theorem due to Emmy Noether, which says that if a system has a continuous symmetry (i.e. if the symmetry holds for infinitesimal transformations) then there is some conserved charge associated to this symmetry.}. 

\subsection{Fitting particles into representations}

Now let us look into some of these representations in detail and interpret them physically in the context of isospin.

Let us first consider the $\j=0$ representation, which is one-dimensional. Because of this dimensionality, we call the state $\ket{\j=0,m=0}$ a \emph{singlet}. Under an isospin rotation it transforms as
\mbox{$
    \ket{0,0}\to U^{\j=0} \ket{0,0},
$}
with $U_{\j=0}$ a one-dimensional unitary matrix of unit determinant, which implies $U^{\j=0}=1$. In other words, a $\j=0$ state is unaffected by an isospin redefinition. This can only mean that this particle carries no isospin at all, and therefore does not interact via the nuclear force. So any particle that is neuter under this force belongs to this isospin representation.

The $\j=1/2$ representation is two-dimensional. It is also called the \emph{fundamental} representation\footnote{
A fundamental representation is one that serves as building blocks for other higher-dimensional representations via tensor product operations. But for unitary groups SU($n$) the \emph{fundamental} representation coincides with the so-called \emph{defining} representation, which consists of the $n$-dimensional vector space where the very definition of these matrices act. For instance, SU($n$) is defined as the group of $n\times n$ unitary matrices, and these matrices can be seen as operators acting on $n$-dimensional vectors, so this $n$-dimensional representation is the defining representation (and also a fundamental) of SU($n$).}, because all the others can be built from tensor products of this space with itself, as we will see in the next section. We have already met particles that fit into this space: the proton and the neutron, corresponding to states $p=\ket{1/2, +1/2}$ and $n=\ket{1/2, -1/2}$.

Moving on to the $\j=1$ representation, we know this is a 3-dimensional space, so the corresponding states $m=-1, 0, 1$ are said to form a \emph{triplet}. If this representation is realized in nature, we would expect to find three particles with near-degenerate masses that take part in the nuclear interaction. Such particles indeed exist: the pions $\pi^{+}$, $\pi^{0}$, and $\pi^{-}$ have masses~\cite{ParticleDataGroup:2022pth}
\begin{equation*}\begin{split}
    m_{\pi^{\pm}} &= 139.57039 \pm 0.00018~\text{MeV},\\
    m_{\pi^0} &= 134.9768 \pm 0.0005~\text{MeV},
\end{split}\end{equation*}
with a relative mass difference
\begin{equation*}
    \frac{m_{\pi^\pm} - m_{\pi^0}}{m_{\pi^\text{avg}}} \sim 3.3\%.
\end{equation*}

Just like the $\j=1/2$ representation, the $\j=1$ representation has a special name. It is called the \textit{adjoint} representation. 
Recall that finding a representation means finding a vector space of states where the generators $\J_i$ act as transformations. 
The $\j=1/2$ representation is a 2-dimensional complex vector space where the generators act as in eq.~\eqref{eq:Js2d}. But the generators $\J_1, \J_2$ and $\J_3$ also form a vector space among themselves, being a basis of the space of traceless hermitean matrices! So what if we use this space, i.e. the algebra itself, as the representation space? We then have a state space constituted of $|\J_1\rangle, |\J_2\rangle$ and $|\J_3\rangle$, and we can define the action of a generator over another via the commutator, i.e.
\begin{equation}
    {\J_i}\ket{\J_j} \equiv [\J_i,\J_j].
    \label{eq:adjoint}
\end{equation}
So, in this representation, the generators will act both as vectors and as operators acting on these vectors. For example, we can use the explicit form of the commutation relations~\eqref{eq:J3pm_algebra} to define the matrix of the $\J_3$ operator acting on $\{\J_-, \J_3, \J_+\}$ as
\begin{equation}
    \J_3 = \begin{pmatrix} -1 & & \\ & 0 & \\ & & +1\end{pmatrix}.
    \label{eq:J3adjoint}
\end{equation}
We can associate $\ket{\J_-}, \ket{\J_3}$ and $\ket{\J_+}$ with the $\pi^-, \pi^0$ and $\pi^+$, respectively.

Naturally, every Lie algebra (i.e. the algebra of infinitesimal transformations related to an arbitrary continuous group) has the adjoint representation. We will see another example in section~\ref{sec:repsSU3} below.

\subsection{Charges, isospin and baryon number}

Looking at the matrix form of $\J_3$ in the adjoint representation, eq.~\eqref{eq:J3adjoint}, one cannot help but noticing that its eigenvalues are precisely the electric charges of the pions (which are the particles belonging in this adjoint representation). We can thus say that, in the adjoint representation, the charge operator is
\begin{equation*}
    Q_\pi = \J_3.
\end{equation*}

If we try to apply this same rule to the fundamental representation, the outcome would be $ Q\,p = +\frac12 p$ and $Q\,n = -\frac12 n$, meaning the proton and neutron would be eigenstates of charge with eigenvalues $\pm 1/2$. This is clearly wrong. However, this problem can be remedied by defining a new quantum number $B$, called baryon number, such that
\begin{equation}
    Q = \J_3 + \frac{B}{2}.
    \label{eq:Q}
\end{equation}
Then the correct charges are found by attributing $B=1$ for protons and neutrons, and $B=0$ for pions. Thus we say that nucleons are \emph{baryons} (i.e. have baryonic number), whereas pions belong to another class of particles called \emph{mesons}.

It turns out that this baryon number $B$ has a deeper physical meaning than the above argument make it seems. More than being a mere adjustment factor to the charge formula, this number is actually a conserved charge of all nuclear reactions. The reader can check that this is indeed the case for all the reactions discussed henceforth in this paper.

\section{Sum of isospin and Clebsch-Gordan coefficients}
\label{sec:sum}

Consider now a system composed of two particles, each carrying some total isospin $\j_1$ and $\j_2$. The composite system must also belong to some SU(2) representation, say, one with total isospin $\J$. The question of which values of $\J$ are allowed for given $\j_1$ and $\j_2$ is known as the problem of how to sum isospin (or any quantity that behaves as angular momentum) in quantum mechanics.
 
\subsection{General formalism}
\label{sec:CG}

Let us then consider a system composed of two particles, one at a state $|\j_1, m_1\rangle$ and the other at a state $|\j_2, m_2\rangle$. The description of this composite system is given by the tensor product 
\begin{equation*}
    |\j_1, m_1\rangle \otimes |\j_2, m_2\rangle \equiv \ket{m_1,m_2}_{\j_1\j_2}.
\end{equation*}
Our aim is to determine how this state can be written as a (superposition of) $|\I, M\rangle$.

First note that an isospin transformation acts on the tensor product by acting on each state separately with the corresponding submatrices $U^\j$ of eq.~\eqref{eq:bigU}, i.e.
\begin{widetext}
    \begin{equation} 
    U^{\j_1\otimes \j_2}\ket{\j_1,m_1}\otimes\ket{\j_2,m_2}
    = 
        U^{\j_1} \ket{\j_1,m_1}\otimes U^{\j_2}\ket{\j_2, m_2}.
    \label{eq:Ux}
    \end{equation}
\end{widetext}
In other words, an isospin transformation on the composite system is obtained by transforming the isospin of the components separately, and then composing (i.e. taking the tensor product) of the transformed states. For infinitesimal transformations, $U^\j = \mathds{1} + i\vec{\theta}\cdot \vec{\J}^\j$ (and similarly for $U^{\j\otimes \j^\prime}$) and we can show that
\begin{equation}\begin{split}
    \vec{\J}^{\j_1\otimes \j_2}
    \ket{m_1,m_2}_{\j_1\j_2} 
    &= \vec{\J}^{\j_1}\ket{\j_1,m_1}\otimes\ket{\j_2,m_2}\\ &+ \ket{\j_1,m_1}\otimes \vec{\J}^{\j_2}\ket{\j_2,m_2}.
    \label{eq:sumJi}
\end{split}\end{equation}
This means that the isospin components of the composite system is the sum of the components of each particle's. In particular, taking only the $\J_3$ component, we see that 
\begin{equation*}
    \J_3^{\j_1\otimes \j_2}
    \ket{m_1, m_2}_{\j_1, \j_2}
    = (m_1+m_2)\ket{m_1,m_2}_{\j_1\j_2},
\end{equation*}
meaning $M=m_1+m_2$.

What about the allowed values of $\I$, the total isospin of the composite system? First note that the maximum value of $M$ is $\j_1+\j_2$, and this must correspond to the maximum value of $\I$ (recall that in the $\I$ representation $M$ varies from $-\I$ to $+\I$, and if there is a $\I>\j_1+\j_2$ there would be a state with $M>\j_1+\j_2$). Moreover, in general $\I$ could vary by half steps, as in eq.~\eqref{eq:js}. But if $\I$ is integer (resp. half-integer) then all allowed values of $M$ are integers (resp. half-integers). But $M$ is fixed as $m_1+m_2$, so we know it to be integer or half-integer, and then $\I$ must have the same characteristic. So $\I$ must vary in integer steps. Knowing $\I_\text{max}=\j_1+\j_2$, and that it varies in integer steps, we can determine the minimum $\I_\text{min}$ by counting the total number of states available. Of course, the counting must match whether we do it in the $\ket{\j_1,m_1}\otimes\ket{\j_2, m_2}$ notation or in the $\ket{\I,M}$ notation. Since one particle can occupy $2\j_1+1$ states, and the other has $2\j_2 +1$ states, we have in total $(2\j_1+1)(2\j_2+1)$ states for the composite system. But for each $\I$ the total number of states is $2\I+1$, so
\begin{equation*}
    \sum_{\I=\I_\text{min}}^{\I_\text{max}} (2\I+1) = (2\j_1+1)(2\j_2+1).
\end{equation*}
It follows that $\I_\text{min} = |\j_1-\j_2|$.

So the state space of a composite system of particles with isospin $\j_1$ and $\j_2$ can be written as states $\ket{\I,M}$, with the total angular momentum $\I$ and the component $M$ satisfying the rules
\begin{enumerate}[label=(\roman*)]
    \item\label{sumJ} $|\j_1-\j_2|\leq \I \leq \j_1+\j_2$,
    \item\label{sumMJ} $M = -\I, ..., \I$ (varying in unit steps),
    \item\label{sumM} $M = m_1 + m_2$.
\end{enumerate}

Notice, in passing, how the adjoint representation $\I=1$ can be obtained from a tensor product of two fundamental ($\j=1/2$) representations. In fact any semi-integer $\I$ can be obtained from tensor products of multiple $\j=1/2$ spaces. It is in this sense that we call the $\j=1/2$ the \emph{fundamental} representation, since it acts as a building block for every other.

\subsection{Clebsch-Gordan coefficients}

We can do better than this: we can find the coefficients of the expansion of $\ket{\I,M}$ in terms of $\ket{\j_1,m_1}\otimes\ket{\j_2,m_2}\equiv \ket{m_1,m_2}_{\j_1\j_2}$ using a constructive method. Let us exemplify it by treating a composite system of $\j_1=1/2$ and $\j_2=1$ (say, a two-particle state of a nucleon and a pion). We will show a physical application of this situation in the next section.

Condition~\ref{sumJ} tells us that the total angular momentum is $\I=3/2$ or $\I=1/2$. From condition~\ref{sumMJ} the state $\ket{\frac12,\frac12}\otimes \ket{1,1}$ (i.e. a proton and a $\pi^+$) corresponds to $M=3/2$, and this state is only accessible in the representation $\I=3/2$, so
\begin{equation*}
    \left|\frac32, \frac32\right\rangle_{\I,M} = 
    \left|\frac12,1\right\rangle_{\frac12,1},
\end{equation*}
where the subscript $\I,M$ clarifies that this state is written in terms of the total isospin and the subscript on the right-hand side tells us the total isospin of each particle.

Now apply the ``lowering'' operator $\J^{\j_1\otimes \j_2}_{-}$ to the state on the left-hand side. Using eqs.~\eqref{eq:sumJi} and~\eqref{eq:cpm} we find
\begin{equation*}
    \left|\frac32,\frac12\right\rangle_{\I,M}
    = \sqrt{\frac13}\left|-\frac12,1 \right\rangle_{\frac12,1}
    + \sqrt{\frac23}\left|\frac12, 0\right\rangle_{\frac12, 1}.
    \label{eq:32_12}
\end{equation*}

Applying the ``lowering'' operator once again gives
\begin{equation*}
    \left|\frac32,-\frac12\right\rangle_{\I,M}
    = \sqrt{\frac23}\left|-\frac12,0\right\rangle_{\frac12,1}
    + \sqrt{\frac13}\left|\frac12,-1\right\rangle_{\frac12,1}.
    \label{eq:32_-12}
\end{equation*}
Finally, lowering once again concludes the $\I=3/2$ representation with
\begin{equation*}
    \left|\frac32, -\frac32\right\rangle_{\I,M}
    = \left|-\frac12,-1\right\rangle_{\frac12, 1},
\end{equation*}
which was expected since only the state $m_1=-1/2$ and $m_2=-1$ can yield $M=-3/2$.

What about the $\I=1/2$ states? We know that they belong in a subspace with different $\J^2$ eigenvalue. Since this operator is hermitian, its eigenspaces are mutually orthogonal. The $|\frac12,\frac12\rangle$ state must be a combination of $m_1=1/2$ and $m_2=0$ or $m_1=-1/2$ and $m_2=1$, and this combination must be orthogonal to the state in eq.~\eqref{eq:32_12}. It follows that
\begin{equation*}
    \left|\frac12,\frac12\right\rangle_{\I,M} = 
    \sqrt{\frac23}\left|-\frac12,1\right\rangle_{\frac12, 1}
    - \sqrt{\frac13}\left|\frac12,0\right\rangle_{\frac12,1}.
\end{equation*}
Finally, the state $|\frac12,-\frac12\rangle_{\I,M}$ can be obtained either by applying a lowering operator to the state above, or by finding a state orthogonal to $\left|\frac32,-\frac12\right\rangle_{\I,M}$. Either way, we find
\begin{equation*}
    \left|\frac12,-\frac12\right\rangle_{\I,M} = 
    \sqrt{\frac13}\left|-\frac12,0\right\rangle_{\frac12,1}
    - \sqrt{\frac23}\left|\frac12,-1\right\rangle_{\frac12,1}.
\end{equation*}

What we have done above is to express $|\I,M\rangle$ as linear combinations of the basis $|\j_1,m_1\rangle\otimes |\j_2,m_2$. The coefficients of this expansion are called \emph{Clebsch-Gordan coefficients}, and we will see in the next section one physical prediction that can be derived from them. They can be obtained from this laborious yet straightforward method. In practice, one simply uses tabulated values as can be found e.g. in~\cite{ParticleDataGroup:2022pth}.

\subsection{Quantitative predictions from symmetry arguments}
\label{sec:quantitative}

Now let us put the above machinery to good use in concrete physical applications.

Given a proton $p=|\frac{1}{2}, \frac{1}{2}\rangle$ and a neutron $n=|\frac12, -\frac12\rangle$, one could think of building composite nuclei containing two such nucleons, which would be described by isospin states
\begin{equation*}
    \left.\begin{array}{cl} 
        pp &= |1,1\rangle\\
        \frac{1}{\sqrt{2}}(pn+np) &= |1,0\rangle\\
        nn &= |1,-1\rangle
        \end{array}\right\} ~\parbox{2.2cm}{\centering triplet state\\ total isospin $1$}
\end{equation*}
and
\begin{equation*}
    \frac{1}{\sqrt{2}}(pn - np) = |0,0\rangle~\rightarrow~\text{singlet}.
\end{equation*}
The factors $\pm{1}/{\sqrt{2}}$ are the Clebsch-Gordan coefficients for the isospin sum of $\j_1=\j_2=1/2$, as the reader can easily calculate using the method developed in the previous section.

It turns out that only one such nucleus exists as a stable configuration in nature: the deuteron, composed of one proton and one neutron. A bound state of two protons and two neutrons turns out to be unstable. Now, in the list of states above there are two of them composed of proton+neutron: the $|1,0\rangle$ and the $|0,0\rangle$. Which corresponds to the deuteron? Note that the states $pp$ and $nn$ are obtainable from $|1,0\rangle$ via a isospin rotation, and if this is a symmetry then the three states in the isospin $=1$ triplet would all be physically equivalent. Since $pp$ and $nn$ are not found as stable states in nature, $|1,0\rangle$ can't be either. Thus isospin symmetry establishes the total isospin of the deuteron to be $0$.

Symmetry arguments can also be used to lead to quantitative predictions regarding reaction cross sections. Consider, for instance, the four reactions\footnote{Here the order of reactants is relevant, because it distinguishes the incoming beam from the target of the reaction in the laboratory frame. Thus reactions (b) and (c) are physically distinguishable.} where scattering of nucleons yields a deuteron $d$ and a pion,
\begin{align*}
    &\text{(a)}~p+p\to d+\pi^+,\\
    &\text{(b)}~p+n\to d+\pi^0,\\
    &\text{(c)}~n+p\to d+\pi^0,\\
    &\text{(d)}~n+n\to d+\pi^-.
\end{align*}
Let us call ${H}$ the Hamiltonian operator describing the dynamics of these nuclear interactions. In principle if we know ${H}$ we can determine the amplitude of any such process via
\begin{equation}
    \mathcal{M} = \langle \text{final}| {H} |\text{initial}\rangle.
    \label{eq:M}
\end{equation}
However, we know that the nuclear interaction is rather non-trivial (being a residual of the more fundamental interaction between quarks), so one can expect ${H}$ to have a complicated form. Moreover, we might not know this form beforehand (which was the case when physicists were still developing the theory of the nuclear interaction in the early 1930s). But even without this knowledge we can already make quantitative statements about the cross sections of the reactions above, based only on symmetry!

One way to proceed, in this case, would be to use Shmushkevich's principle~\cite{MacFarlane:1965wp, Wohl:1982}, which follows from isospin symmetry. It states that if the population of reactants is uniformly distributed among all its possible isospin states, then the population of products will also be uniform (according to this same criterion) at all times. For instante, consider a uniform beam of nucleons incident on a uniform target of nucleons. Uniform distribution means that, for each 100 pairs of nucleons on beam/target, half of them will be protons and half neutrons. Each pair will react according to one of the four reactions above, and in the end we will have 100 deuterons and 100 pions. The deuteron is a singlet, so its distribution will be uniform regardless. Now, Shmushkevich's principle states that the pions must be uniformly distributed among $\pi^-, \pi^0, \pi^+$. But there are two channels that can lead to a $\pi^0$, namely (b) and (c), while only (a) can lead to $\pi^+$ and only (d) can lead to $\pi^-$. Thus a uniform distribution of the pions in the final products can only be achieved if (a) and (d) are twice more efficient than (b) and (c). In other words, the cross sections must be in the ratio
\begin{equation}
    \sigma_a:\sigma_b:\sigma_c:\sigma_d = 2:1:1:2.
    \label{eq:sNN}
\end{equation}
This is in very good agreement with experimental results~\cite{Fliagin:1959}.

These results can also be obtained by a more systematic method involving the Clebsch-Gordan coefficients. We will illustrate this for the case of a scattering experiment of charged pions by protons\footnote{We leave it as an exercise to the reader to show that eq.~\eqref{eq:sNN} can also be reached using the method delineated below.}. There are possible reactions, namely
\begin{align*}
    &\text{(a)}~\pi^+ + p\to \pi^+ + p,\\
    &\text{(b)}~\pi^-+p\to \pi^- + p,\\
    &\text{(c)}~\pi^- + p\to \pi^0 + n.
\end{align*}
Each pair of reactants/products can be written as a combination of states with total isospin $3/2$ and $1/2$ by inverting the relations found in section~\ref{sec:sum}. Specifically, recalling that $p=|\frac12,\frac12\rangle$ and $n=|\frac12,-\frac12\rangle$ whereas $\pi^-,\pi^0,\pi^+$ correspond to $|1,-1\rangle, |1,0\rangle$ and $|1,1\rangle$, we have
\begin{align*}
    \pi^+ p &= \left|\frac32,\frac32\right\rangle_{\I,M},\\
    \pi^0 n &= \sqrt{\frac23}\left|\frac32,-\frac12\right\rangle_{\I,M} 
        + \sqrt{\frac13}\left|\frac12,-\frac12\right\rangle_{\I,M},\\
    \pi^- p &= \sqrt{\frac13}\left|\frac32, -\frac12\right\rangle_{\I,M} - \sqrt{\frac23}\left|\frac12, -\frac12\right\rangle_{\I,M}.
\end{align*}

When applying eq.~\eqref{eq:M} we will find terms of the form $\left\langle \I^\prime, M^\prime\right| {H}\left|\I,M\right\rangle$. We have seen in section~\ref{sec:conservation} that if $\I\neq \I^\prime$ these terms vanish due to isospin conservation (which follows from the symmetry). The same reasoning leads to vanishing amplitude if $M\neq M^\prime$. Moreover, the amplitude cannot depend on $M$, because an $M^\prime$ state can be obtained from $M$ via an isospin redefinition, which is a symmetry of the nuclear dynamics and thus cannot affect the physics\footnote{Technically, $|\I,M^\prime\rangle = U|\I,M\rangle$ for some transformation matrix $U$. Then $\langle \I,M^\prime|H|\I,M^\prime\rangle = \langle \I,M|U^\dagger H U|\I,M\rangle = \langle \I,M|H|\I,M\rangle$, since $[H,U]=0$ and $U$ is unitary.}. This means that, from symmetry alone, we can establish that
\[
    \langle \I^\prime, M^\prime| H |\I,M\rangle = \mathcal{M}_\I\, \delta_{\I\I^\prime}\delta_{MM^\prime}.
\]
We therefore expect only two kinds of amplitudes: $\mathcal{M}_{3/2}$ and $\mathcal{M}_{1/2}$. It is as if these processes could proceed via two mediating channels, each corresponding to a different total isospin, and each with a different probability amplitude. Applying eq.~\eqref{eq:M} yields $\mathcal{M}_a = \mathcal{M}_{3/2}$, $\mathcal{M}_b =(\mathcal{M}_{3/2} + 2\mathcal{M}_{1/2})/3$ and $\mathcal{M}_c = \sqrt{2}(\mathcal{M}_{3/2} - \mathcal{M}_{1/2})/3$. Since the cross sections scale with $\sigma\sim |\mathcal{M}|^2$, one finds
\begin{widetext}
    \[\sigma_a: \sigma_b: \sigma_c 
    = 9\left|\mathcal{M}_{\frac32}\right|^2 : \left|\mathcal{M}_{\frac32}+2\mathcal{M}_{\frac12}\right|^2 : 2\left|\mathcal{M}_{\frac32} - \mathcal{M}_{\frac12}\right|^2.
    \]
\end{widetext}

\begin{figure}
    \centering
    \includegraphics[width=.45\textwidth]{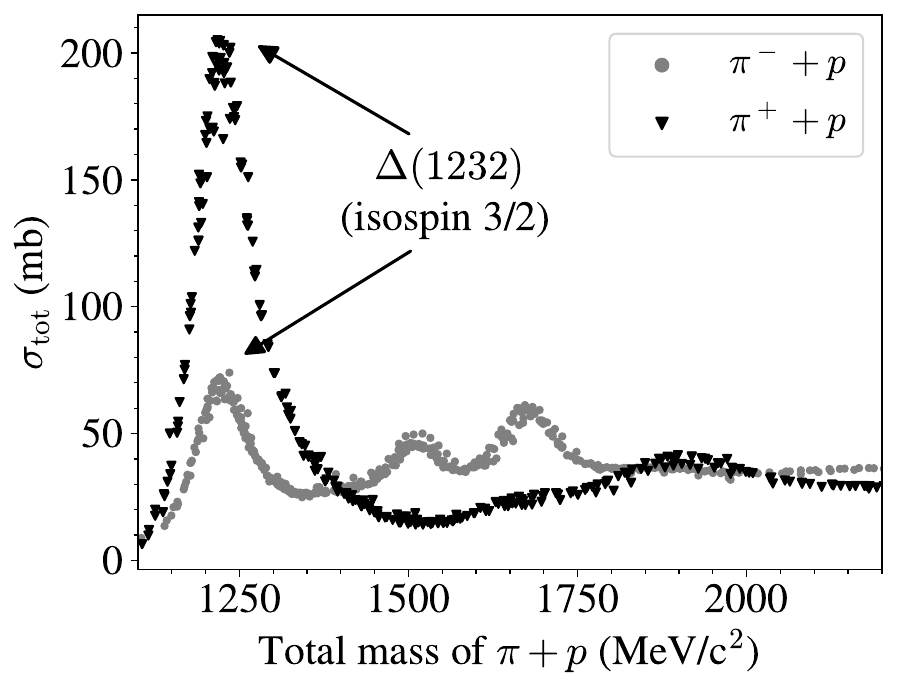}
    \caption{Total cross section for $\pi^+ + p$ (black triangles) and $\pi^- + p$ (gray bullets) as function of the total centre-of-mass energy of the reaction. Data from ref.~\cite{ParticleDataGroup:2022pth}.}
    \label{fig:pip}
\end{figure}

When we perform these collision experiments and measure the cross section, we find certain bumps where the cross section is significantly increased, as shown in figure~\ref{fig:pip}. These are called \emph{resonances} and can be interpreted as a consequence of an intermediate unstable particle mediating the reaction. In figure~\ref{fig:pip} we see a very pronounced peak when the reaction occurs at center-of-mass energy of $\approx 1232$~MeV. This means that, at these energies, there is a new intermediate path contributing to the reaction, namely
\[
    \pi + p \to \Delta^* \to \text{products},
\] 
i.e. the reactants first form a new particle, called a $\Delta$, which then decays to the observed products (the $^*$ over the name of the particle indicates it is unstable). The existence of this particle thus enhances the cross section, leading to the peaks shown in figure~\ref{fig:pip}.

The $\Delta(1232)$ is a particle with total isospin $3/2$, which means that, at these energy scales, the $3/2$ channel is much more significant than the $1/2$, i.e. $\mathcal{M}_{3/2}\gg \mathcal{M}_{1/2}$. Then the total cross sections of $\pi^+ + p$ and $\pi^-+p$ reactions satisfy
\[
    \frac{\sigma_\text{tot}(\pi^+ + p)}{\sigma_\text{tot}(\pi^- + p)} = \frac{\sigma_a}{\sigma_b+\sigma_c}\approx \frac{9}{1+2} = 3.
\]
This is in excellent agreement with the behaviour shown in fig.~\ref{fig:pip}.

Incidentally, since the $\Delta$ belongs in the $\j=3/2$ representation, we expect to find $2\j+1=4$ such particles of approximately degenerate masses. Moreover, noticing that these $\Delta$'s are baryons (since they result from a collision of $\pi+p$, with $B_\Delta = B_\pi + B_p = 1$), we expect from eq.~\eqref{eq:Q} that these four particles would have electric charges $+2, +1, 0, -1$ (in units of elementary charge). This is indeed in agreement with the experiments, which have observed a $\Delta^{++}$, $\Delta^{+}$, $\Delta^0$ and $\Delta^-$.

\section{Extending the symmetry}
\label{sec:su3}

\subsection{Strange particles and the particle zoo}
\label{sec:zoo}

If the world was composed only of protons, neutrons and pions, it seems that an SU(2) isospin symmetry would be an excellent explanation for the classification and behaviour of these particles.

But in the late 1940s and early 1950s physicists were confronted with the discovery of a number of other particles. First, cloud chamber photographs of cosmic rays showed some unusually heavy particles decaying into pions ~\cite{Rochester:1947mi, Brown:1949mj}, evidencing yet unknown mesons which later came to be known as the kaons. In 1950 a neutral baryon decaying into protons and pions was discovered in cosmic ray events, which later came to be dubbed the $\Lambda^0$ baryon~\cite{Hopper:1950}. It did not take long to notice that there was something strange about these particles. For instance, $\Lambda^0 + 2\pi$'s could be produced copiously when colliding proton + pions, from which one can infer the transition amplitude $\langle \Lambda^0 \pi\pi | H | p\pi\rangle$. If one then used this amplitude to estimate the decay lifetime of $\Lambda^0\to p+\pi^-$, one would find $\sim 10^{-23}$~s. However, observations showed that the correct lifetime was $\sim 10^{-10}$~s: a discrepancy of 13 orders of magnitude! Clearly, production and decay were very different mechanisms for $\Lambda^0$ (and the same was true for the kaons).

The way out of the problem involved two steps. At first, it was postulated that proton+pion collisions produce $\Lambda^0$ only in association with kaons,
\begin{align*}
    \pi^- + p \to \Lambda^0
    &+ \underset{|}{K^0}
    \\[-4mm]
    &\quad~\, \to \pi^- + \pi^+.
\end{align*}
This is known as ``associated production'', namely that these ``exotic'' particles would always be produced in pairs. For some time this was enough to explain all observations, but soon new reactions were observed involving an odd number of exotic particles, while other reactions involving pairs were never observed despite being allowed by the conjecture\footnote{The hypothesis of associated production came into crisis when the reaction $\pi^{-}+p\rightarrow\Xi^{-}+K^{0}+K^{+}$ was observed (with $\Xi^-$ an exotic baryon heavier than the proton), while the reaction $\pi^{-}+p\rightarrow\Xi^{-}+K^{+}$, allowed by the conjecture, was not. These can be explained by assigning to $\Xi^-$ a strangeness value of $S=-2$ and assuming that strangeness is an approximately conserved charge.}. Later it was recognized that there was another (approximately valid) conservation law involved: one can assign to these new particles a new quantum charge, which was (appropriately enough) called \emph{strangeness}, such that $S_{\Lambda^0}=-1$ and $S_{K^0}=+1$. This way, the production channel above conserves strangeness ($S_\text{before}=S_\text{after}=0$), whereas the decays $\Lambda^0\to p+\pi^-$ and $K^0\to \pi^-+\pi^+$ do not. This explains why the decays are much less favourable than the production. 

Now, if we were to fit these strange particles in isospin representations, we would be forced to place the $\Lambda^0$ in a singlet state, since no other particle looks like an SU(2) partner\footnote{The discovery of other strange particles called $\Sigma^\pm$ could, upon first sight, lead to a tentative classification of $(\Sigma^-,\Lambda^0, \Sigma^+)$ as an isospin triplet. However, soon a new strange particle $\Sigma^0$ was discovered through the decay $\Sigma^0\to \Lambda^0+\gamma$, with a mass that was much closer to $\Sigma^\pm$. So the true isospin triplet is composed of $(\Sigma^-, \Sigma^0, \Sigma^+)$, leaving $\Lambda^0$ as a singlet.} of $\Lambda^0$. In this case the expression~\eqref{eq:Q} for the charge operator would again fail to give the correct charge of $\Lambda^0$, forcing us to modify it as
\begin{equation}
    Q= T_3 + \frac{B+S}{2}.
    \label{eq:GMNishijima}
\end{equation}

If we now try to assign quantum numbers to the kaons to meet their correct charges, we would be led to assign $T_3=-1/2$ to $K^0$ (to match $Q_{K^0}=0$ since $B_{K^0}=0$ and $S_{K^0}=1$), so that the $T_3=+1/2$ partner would be a $K^+$. The antiparticle of $K^+$ would be a $K^-$, with opposite strangeness charge $S=-1$, so it cannot belong to the same isospin multiplet as the latter (contrary to charge, strangeness is independent from isospin, and particles in the same isospin multiplet have the same strangeness). It follows that there must be another isospin $1/2$ multiplet where we fit the $K^+$ and another neutral kaon which we call $\overline{K^0}$, which is the antiparticle of $K^0$ and must therefore have the same mass. This particle was predicted this way, and later indeed observed~\cite{Bardon:1958guy}, confirming the success of this scheme based on strangeness.

By the late 1960s a number of additional heavy strange particles were discovered, such as $\Sigma$ and $\Xi$ baryons, as well as the (non-strange) $\Delta$'s already mentioned above. Due to this proliferation of particles, a classification based solely on SU(2) representations became increasingly more cumbersome, since additional representations (often with the same isospin as others, yet fitting rather different particles) had to be introduced \emph{ad hoc} to make the scheme work. This proliferation of particles and the difficulty in classifying them neatly under SU(2) representations is colloquially known as the \emph{particle zoo}.

\subsection{Sakata model and SU(3) symmetry}
\label{sec:sakata}

A major step towards a more elegant and more predictive classification of hadrons was made by japanese physicist Shoichi Sakata in 1956. Sakata proposed that the $\Lambda^0$ would be yet another state of a nucleon, which could then be found in three different states,
\begin{align}
    p^+ = \begin{pmatrix}
        1 \\ 0 \\ 0
    \end{pmatrix},
    \quad
    n^0 = \begin{pmatrix}
        0 \\ 1 \\ 0
    \end{pmatrix},
    \quad
    \Lambda^0 = \begin{pmatrix}
        0 \\ 0 \\ 1
    \end{pmatrix},
    \label{eq:su3fundamental}
\end{align}
or rather in a general superposition of these states. Sakata's statement is that the nuclear interaction has an approximate symmetry under a redefinition of these states. So, similarly to what we discussed in section~\ref{sec:isospinsu2}, one could do a redefinition
\begin{equation*}
    N = \begin{pmatrix}
        \psi_p \\ \psi_n \\ \psi_\Lambda
    \end{pmatrix} \to N^\prime= UN,
\end{equation*}
with $U$ a $3\times 3$ unitary matrix with unit determinant. The group of such matrices is called SU(3).

Notice that
\[
    m_{\Lambda^0} = 1115.683\pm 0.006~\text{MeV},
\]
so that the relative mass difference between the $\Lambda^0$ and the average mass of proton and neutron is 
\[
    \frac{\Delta m_{\Lambda^0,pn}}{m_{\Lambda^0,pn}^\text{avg}} \simeq 17.2\%.
\]
This already illustrates that this new symmetry is only approximately valid, but this shall not prevent us from proceeding. After all, SU(2) isospin was also only approximate but still gave us many important and valid predictions.

~\\
\subsection{Representations of SU(3)}
\label{sec:repsSU3}

Our task is again to build representations of this SU(3) group, and then argue that each representation corresponds to a certain particle type. The method is the same as above: we build representations by performing infinitesimal SU(3) transformations on a certain reference state. Following the same arguments as in section~\ref{sec:su2inftl} we find that the generators of infinitesimal SU(3) transformations must be hermitean traceless matrices.

We already know one representation of this group: the fundamental representation, a 3-dimensional space with basis given by eq.~\eqref{eq:su3fundamental}. In this representation, the infinitesimal transformations are given by $3\times 3$ complex traceless hermitian matrices. How many generators are there? As a real vector space, the space of all complex $3\times 3$ matrices has dimension\footnote{There are $3^2$ entries in the matrix, and each is described by two real numbers, namely the real and the imaginary parts of each complex entry.} $2\times3^2$. Making it hermitian decreases the dimension in half\footnote{The $3$ diagonal elements must be real. Once we fix the $2\times 3$ elements in the upper-triangular matrix (above the diagonal), those below the diagonal are also fixed by the hermiticity condition. So we only need $3+2\times 3=9$ real numbers to describe the whole matrix.}, leaving $3^2$. Finally the traceless condition lowers the dimension once more by 1, leaving us with a dimension\footnote{In general, the Lie algebra of $\mathfrak{su}(n)$ (i.e. the space of $n\times n$ hermitean traceless matrices) has dimension $n^2-1$. Applying this to $\mathfrak{su}(2)$ one finds a dimension $2^2-1=3$, as we saw in section~\ref{sec:su2inftl}.} of $3^2-1$. This means SU(3) has 8 linearly independent generators. In SU(2), the usual generators are given by the Pauli Matrices. We also have, for 3 dimensions, an usual choice of generators, called the {Gell-Mann} matrices, given by
\begin{widetext}
\begin{equation}\begin{split}
    \lambda_1 = \begin{pmatrix}
          0 & 1 & 0\\
          1 & 0 & 0\\
          0 & 0 & 0
       \end{pmatrix},\quad
       \lambda_2 &= \begin{pmatrix}
          0 & -i & 0\\
          i & 0 & 0\\
          0 & 0 & 0
       \end{pmatrix},\quad
       \lambda_3 = \begin{pmatrix}
          1 & 0 & 0\\
          0 & -1 & 0\\
          0 & 0 & 0
       \end{pmatrix},\\
       \lambda_4 = \begin{pmatrix}
          0 & 0 & 1\\
          0 & 0 & 0\\
          1 & 0 & 0
       \end{pmatrix},\quad  
        \lambda_5 &= \begin{pmatrix}
          0 & 0 & -i\\
          0 & 0 & 0\\
          i & 0 & 0
       \end{pmatrix},\\ 
       \lambda_6 = \begin{pmatrix}
          0 & 0 & 0\\
          0 & 0 & 1\\
          0 & 1 & 0
       \end{pmatrix},\quad 
        \lambda_7 &= \begin{pmatrix}
          0 & 0 & 0\\
          0 & 0 & -i\\
          0 & i & 0
       \end{pmatrix},\quad 
       \lambda_8 = \dfrac{1}{\sqrt{3}}\begin{pmatrix}
          1 & 0 & 0\\
          0 & 1 & 0\\
          0 & 0 & -2
       \end{pmatrix}. 
\end{split}\end{equation}
\end{widetext}
Notice that $\lambda_{1,2,3}$ are simply the Pauli matrices inserted into a larger $3\times 3$ paradigm. This means that the algebra $\mathfrak{su}(2)$ is contained in $\mathfrak{su}(3)$. Consequently, representations of SU(3) contain representations of SU(2). In other words, the classification scheme we will construct here will extend what we discussed above: all the successes of classification by SU(2) isospin will be maintained. 

Actually, a glimpse at $\lambda_4$ and $\lambda_5$ shows that, if we remove the second row and column, we are left with two of the Pauli matrices. And the third Pauli matrix would be obtained from $(\lambda_3 + \sqrt{3}\lambda_8)/2$. Likewise, $\lambda_6$ and $\lambda_7$ also contain the Pauli matrices as sub-blocks (try removing the first row and first column of these matrices and see what is left). The third Pauli matrix in this sub-block is obtained from $(-\lambda_3 + \sqrt{3}\lambda_8)/2$. So we see that the $\mathfrak{su}(3)$ algebra is formed of three overlapping $\mathfrak{su}(2)$~\cite{Zee:2016fuk}. They are overlapping because the third Pauli matrix of these subalgebras are not mutually independent.

We will apply the same argument as we did for $\mathfrak{su}(2)$ above, and the same construction method, to obtain representations of $\mathfrak{su}(3)$. First, notice that $T_3$ and $Y \equiv \frac{2}{\sqrt{3}}T_8$ are commuting hermitian generators. This means they can be simultaneously diagonalized, so the states of an SU(3) representation will be labeled as\footnote{This is analogous to the states of an SU(2) representation being labeled by ${m}$, the eigenvalue of $\J_3$. In that case $\j$ labeled the representation itself. To avoid notation cluttering, for SU(3) we will not include the label inside the kets denoting the representation.}
$ \ket{t_3, y}$, with $t_3$ and $y$ eigenvalues of $T_3$ and $Y$, respectively.

Furthermore, since we have three overlapping $\mathfrak{su}(2)$ subalgebras, we can define three sets of raising and lowering operators, namely
\begin{equation}\begin{split}
    T_{\pm}=T_1\pm iT_2,\\
    U_{\pm}=T_{6}\pm iT_7,\\
    V_{\pm}=T_4\pm iT_5.
\end{split}\end{equation}
In terms of these operators, one can write the following important commutation relations:
\begin{align*}
    [T_3,T_\pm]&=\pm T_{\pm},
    \qquad &&[Y,T_\pm]=0,\\
    [T_3,V_\pm]&=\pm\frac{1}{2}V_{\pm},
    \qquad &&[Y,V_\pm]=\pm V_{\pm},\\
    [T_3,U_\pm]&=\mp\frac{1}{2}U_{\pm},
    \qquad &&[Y,U_\pm]=\pm U_{\pm},
\end{align*}
\begin{align}
    [T_+, T_-] &= 2T_3, \nonumber\\
    [V_+, V_-] &= T_3 + \frac{3}{2}Y \equiv 2V_3,
    \label{eq:su3algebra}
    \\
    [U_+, U_-] &= -T_3 + \frac{3}{2}Y \equiv 2U_3, \nonumber
\end{align}
\begin{align*}
    [T_+, V_+] &= 0,\qquad &&[T_+, V_-] = -U_-,\\
    [T_+, U_-] &= 0,\qquad &&[T_+, U_+] = V_+,\\
    [U_+, V_+] &= 0,\qquad &&[U_+, V_-] = T_-,\\
    [T_3, Y] &= 0.
\end{align*}

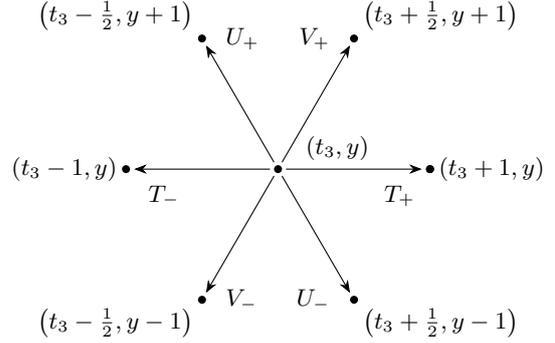
\begin{figure}
    \centering
    \begin{tikzpicture}[scale=0.5]
      \pgfmathsetmacro{\latSize}{4}
      \foreach \i in {-1,...,0} {
        \foreach \j in {-1,...,1} {
          \pgfmathsetmacro{\x}{\latSize*(\i+0.5*Mod(\j,2))}
          \pgfmathsetmacro{\y}{\latSize*sin(60)*\j}
          \fill ({\x}, {\y}) circle (3pt);
        }
        \fill (\latSize,0) circle (3pt);
      }
      % Draw arrow from (0,0) to (2,2) with labels
      \pgfmathsetmacro{\cosine}{\latSize*cos(60)};
      \pgfmathsetmacro{\sine}{\latSize*sin(60)};
      \coordinate (Tp) at (\latSize, 0);
      \coordinate (Vp) at (\cosine, \sine);
      \coordinate (Up) at (-\cosine, \sine);
      \coordinate (Tm) at (-\latSize, 0);
      \coordinate (Vm) at (-\cosine, -\sine);
      \coordinate (Um) at (\cosine, -\sine);
      
      \draw[-{Stealth}, shorten <=3pt, shorten >=3pt] (0,0) -- node[xshift=-4mm, yshift=-1mm, below, at end] {$T_+$} (Tp);
      \node[right] at (Tp) {\small$\left(t_3+1,y\right)$};
      \draw[-Stealth, shorten <=3pt, shorten >=3pt] (0,0) -- node[xshift=-2mm, left, at end] {$V_+$} (Vp);
      \node[above right] at (Vp) {\small$\left(t_3+\frac{1}{2},y+1\right)$};
      \draw[-Stealth, shorten <=3pt, shorten >=3pt] (0,0) -- node[xshift=2mm, right, at end] {$U_+$} (Up);
      \node[above left] at (Up) {\small$\left(t_3-\frac{1}{2},y+1\right)$};
      \draw[-Stealth, shorten <=3pt, shorten >=3pt] (0,0) -- node[xshift=5mm, yshift=-1mm, below, at end] {$T_-$} (Tm);
      \node[left] at (Tm) {\small$\left(t_3-1,y\right)$};
      \draw[-Stealth, shorten <=3pt, shorten >=3pt] (0,0) -- node[xshift=2mm, right, at end] {$V_-$} (Vm);
      \node[below left] at (Vm) {\small$\left(t_3-\frac{1}{2},y-1\right)$};
      \draw[-Stealth, shorten <=3pt, shorten >=3pt] (0,0) -- node[xshift=-2mm, left, at end] {$U_-$} (Um);
      \node[below right] at (Um) {\small$\left(t_3+\frac{1}{2},y-1\right)$};

      \node[above right] at (0.5,0) {\small $(t_3,y)$};
    \end{tikzpicture}
    
    \caption{The ladder operators $T_\pm$, $U_\pm$ and $V_\pm$ map a point in the $(t_3, y)$ plane to other points as shown in the diagram above. These vectors (and sometimes the ladder operators themselves) are called the \emph{roots} of the $\mathfrak{su}(3)$ algebra.}
    \label{fig:roots}
\end{figure}

To understand these relations, consider for instance the action of the $V_+$ operator on a state $\ket{t_3,y}$. From the above commutation relations we find e.g. $(T_3V_+ - V_+T_3)\ket{t_3,y} = T_3V_+\ket{t_3,y} - t_3V_+\ket{t_3,y} = \frac{1}{2}\ket{t_3,y}$, so that $T_3V_+\ket{t_3,y} = \left(t_3 + \frac{1}{2}\right)V_+\ket{t_3,y}$. In other words, $V_+\ket{t_3,y}$ is an eigenvector of $T_3$ with eigenvalue ($t_3+1/2$). Similarly, from its other commutation relation, it is an eigenvector of $Y$ with eigenvalue $(y+1)$. Therefore all ladder operators map eigenvectors of $(T_3,Y)$ to other such eigenvectors (unless they map $\ket{t_3,y}$ to the zero vector). 

But to act on a state with these ladders is equivalent to performing arbitrary SU(3) transformations on this state. Thus we will build irreducible representations of SU(3) by repeatedly applying these operators on a reference state of the representation, resulting in a number of linearly independent states, until this iterative procedure exhausts itself and no new independent vectors are produced. As long as we are dealing with finite dimensional representations, this iteration must always cease to produce new independent vectors at some point. When this happens, we have found our representation: all states are related to each other by SU(3) transformations. 

In the case of SU(2), only one generator $\J_3$ could be simultaneously diagonalized\footnote{The operator $\J^2$ is not one of the generators, although it is built from them. It is called the \emph{Casimir operator} of the $\mathfrak{su}(2)$ algebra.}, whose eigenvalues $m$ label the states of the representation, and the other generators formed ladder operators, which move the states along both directions of this straight line (increasing and decreasing $m$). Thus SU(2) representations could be visualized as a series of points on a straight line, corresponding to the allowed values of $m=-\j, \ldots, \j$. In the case of SU(3) each state in a representation space is labeled by two parameters, and we can illustrate this representation as points in a $(t_3, y)$ plane. Each point is called a ``weight'' and the collection of points belonging to a representation is called a ``weight diagram''. Figure~\ref{fig:roots} shows how each point (i.e. each weight) is mapped onto another by the ladder operators defined above. 

\subsubsection*{The fundamental representation} 

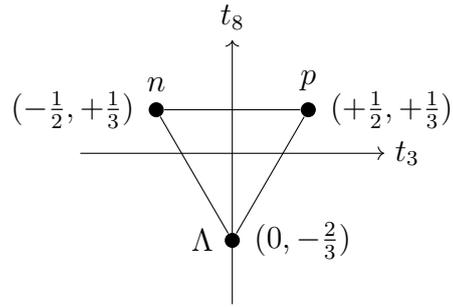
\begin{figure}
    
\begin{tikzpicture}[scale=.5, dot/.style={circle,fill=black, inner sep=2pt}]
       % Eixos
        \pgfmathsetmacro{\latSize}{4}
        \large
        \draw[->] (-\latSize,0) -- (\latSize,0) node[right] {$t_3$};
        \draw[->] (0,-\latSize) -- (0,.75*\latSize) node[above] {$t_8$};
        \node[dot, label=above:${p}$] (p) at (0.5*\latSize,  {\latSize*(sqrt(3)/6)}){};
        \node[dot, label=above:${n}$] (n) at (-0.5*\latSize, {\latSize*(sqrt(3)/6)}){};
        \node[dot, label=left:${\Lambda}$] (L) at (0, {\latSize*(-sqrt(3)/3)}){};

        \node[label=right:{$(+\frac12,+\frac13)$}] at (p){};
        \node[label=left:{$(-\frac12,+\frac13)$}] at (n){};
        \node[label=right:{$(0,-\frac23)$}] at (L){};
        % Triângulo equilátero
        \draw (p) -- (n) -- (L) -- (p);
\end{tikzpicture}
    \caption{Weight diagram of the fundamental representation of SU(3).}
    \label{fig:fundRepresentation}
\end{figure}

Let us consider first the fundamental or defining representation of SU(3), which is the representation where the transformations act as $3\times 3$ unitary matrices (and infinitesimal SU(3) transformations are represented by the Gell-Mann matrices themselves). Going along with the Sakata Model of SU(3), we can label the three vectors belonging to this representation space as $p = (1,0,0), n=(0,1,0), \Lambda=(0,0,1)$. We can easily see that they are eigenvectors of $T_3$ and $Y$ with respective eigenvalues $(-1/2, +1/3)$, $(+1/2,+1/3)$ and $(0,-2/3)$. We can write that as 
\begin{align*}
    n &= \left|-\frac12,+\frac13\right\rangle,\qquad p = \left|+\frac12,+\frac13\right\rangle,\\[3mm]
    &\qquad\qquad\quad\Lambda = \left|0,-\frac23\right\rangle
\end{align*}
and represent it in a $(t_3, y)$ plane as in figure \ref{fig:fundRepresentation}. Since it is three dimensional, we usually refer to it as the \textbf{3} representation.

\subsubsection*{The antifundamental representation}

It turns out that, in SU(3), there is another 3-dimensional representation which is inequivalent to the fundamental $\mathbf{3}$. Note that, if $U$ is an SU(3) matrix, then so is its complex conjugate $U^*$. The $U^*$'s defines the complex-conjugate representation of the $U$'s. Moreover, for an infinitesimal SU(3) transformation\footnote{Recall that the parameters of our transformation (i.e. the ``angles'' by which we perform the ``rotation''/transformation, dubbed $\alpha^a$ below, are always real numbers.}
\begin{equation*}
    U = \mathds{1} + i \alpha^a T_a 
    \implies
    U^* = \mathds{1} + i \alpha^a (-T_a^*).
\end{equation*}
Thus we see that, if the Gell-Mann matrices $\lambda_a$ represent infinitesimal transformations acting on the fundamental representation $p,n,\Lambda$, then the matrices $-\lambda_a^*$ represent transformations on the \emph{antifundamental} (i.e. the complex conjugate of the fundamental).

The reader may now pause and wonder: why was this not mentioned in the case of SU(2) above? Where is the ``antifundamental'' of SU(2)? It turns out that, for SU(2), any representation is equivalent to its conjugate! This is not hard to see: the generators of SU(2) are such that
\begin{equation*}
    -\J_1^* = -\J_1, -\J_2^* = \J_2 \quad\text{and}\quad -\J_3^* = -\J_3.
\end{equation*}
So going from a $\j$ representation to its complex conjugate in SU(2) amounts to mapping $m\to -m$ and $\J_\pm \to -\J_\mp$. Graphically, all states of an SU(2) representation lie along a line symmetric around the origin. Thus the above transformations (going from the fundamental to the antifundamental) correspond to a mere renaming of the states of the $\j$ representation, which keeps the vector space itself unaltered.

\begin{figure}
    \begin{tikzpicture}[scale=.5, dot/.style={circle,fill=black, inner sep=2pt}]
       % Eixos
        \pgfmathsetmacro{\latSize}{4}
        \large
        \draw[->] (-\latSize,0) -- (\latSize,0) node[right] {$y$};
        \draw[->] (0,-.75*\latSize) -- (0,\latSize) node[above] {$t_8$};
        \node[dot, label=below:$\overline{p}$] (pbar) at (-0.5*\latSize,  {\latSize*(-sqrt(3)/6)}){};
        \node[dot, label=below:$\overline{n}$] (nbar) at (0.5*\latSize, {\latSize*(-sqrt(3)/6)}){};
        \node[dot, label=above left:$\overline{\Lambda}$] (Lbar) at (0, {\latSize*(sqrt(3)/3)}){};

        \node[label=left:{$(-\frac12,-\frac13)$}] at (pbar){};
        \node[label=right:{$(+\frac12,-\frac13)$}] at (nbar){};
        \node[label=right:{$(0,+\frac23)$}] at (Lbar){};
        % Triângulo equilátero
        \draw (pbar) -- (nbar) -- (Lbar) -- (pbar);
\end{tikzpicture}
    \caption{Weight diagram for the antifundamental representation of SU(3). Notice how this space is inequivalent to the fundamental representation defined in figure~\ref{fig:fundRepresentation}. In that case the state with highest $t_3$ value was $p$, and we obtained a $\Lambda$ by applying a ladder operator $V_-$ to it. Here, the state with highest $t_3$ is $p^*$, which is annihilated by the action of $V_\pm$.}
    \label{fig:AfundRepresentation}
\end{figure}
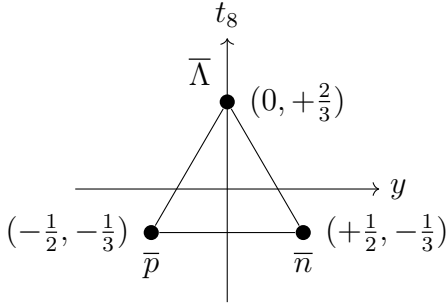

But for SU(3) these two representations define different spaces. If we define a basis $p^*=(1,0,0)$, $n^*=(0,1,0)$ and $\Lambda^*=(0,0,1)$ in the complex conjugate space of the fundamental representation, then the infinitesimal transformations act on these as $T_a^\text{antifund.}\equiv -T_a^*$. In particular, these states are labeled by eigenvalues of $-T_3^*$ and $-Y^*$. Since $T_3$ and $Y$ are real matrices, the labels of $\overline{p}$, $\overline{n}$ and $\overline{\Lambda}$ are just those of $p,n,\Lambda$ with opposite signs\footnote{Just act with $-T_3^*$ and $-Y^*$ on $p^*,n^*$ and $\Lambda^*$ defined above and see what you get.},
\begin{align*}
    &\qquad\qquad\quad \overline{\Lambda} = \left|0,\frac23\right\rangle,\\[3mm]
    \overline{p} &= \left|-\frac12,-\frac13\right\rangle,\qquad \overline{n} = \left|+\frac12,-\frac13\right\rangle.
\end{align*}
The weight diagram in the $(t_3, y)$ plane is depicted in figure~\ref{fig:AfundRepresentation}. 
The antifundamental representation is clearly also 3-dimensional, but different from the fundamental: thus we denote it by $\boldsymbol{\overline{3}}$. 

\subsubsection*{The adjoint representation}

We have mentioned before that every Lie algebra has an adjoint representation, where the operators are seen at the same time as transformations acting on states, and as the very states being acted upon. This action is defined via the commutation relations, as in eq.~\eqref{eq:adjoint}. This means that the adjoint representation of SU(3) is an 8-dimensional vector space with basis $\{\ket{T_3},\ket{Y},\ket{T_\pm},\ket{U_\pm},\ket{V_\pm}\}$. These are all eigenvalues of $T_3$ and $Y$, as one can see for instance from
\begin{align*}
        {T}_3\ket{T_\pm} &= [T_3,T_\pm]=\pm \ket{T_{\pm}}, \\
        {Y}\ket{T_\pm} &= [Y,T_\pm]=0 \ket{T_\pm},
\end{align*}
or 
\begin{equation*}
    (T_3, Y)\ket{T_\pm} = \left(\pm 1, 0\right)\ket{T_\pm}.
\end{equation*}
Likewise the commutation relations of these ladder operators with $T_3$ and $Y$ imply that
\begin{align*}
        ({T}_3, Y)\ket{U_\pm} &= \left(\mp\frac{1}{2}, \pm 1\right)\ket{U_{\pm}}\\
        (T_3, Y)\ket{V_\pm} &= \left(\pm\frac{1}{2}, \pm1\right)\ket{V_{\pm}}\\
        ({T}_3, Y)\ket{T_3} &= \left(0, 0\right) \ket{T_3}\\
        ({T}_3, Y)\ket{Y} &= \left(0,0\right) \ket{Y}.
\end{align*}

\begin{figure}
    \centering
    \begin{tikzpicture}[scale=0.5, dot/.style={circle,fill=black, inner sep=2pt}]
        \normalsize
        \pgfmathsetmacro{\latSize}{4}
        % DRAW THE AXES
        \draw[->] (0, -1.2*\latSize) -- node[above, at end] {$y$} (0, 1.2*\latSize);
        \draw[->] (-1.2*\latSize, 0) -- node[right, at end] {$t_3$} (1.2*\latSize, 0);
        % DEFINE THE COORDINATES OF THE POINTS IN THE ADJOINT REP (up, down, strange and their anti-particles)
        \node[dot, label=above:{$\mathbf{\ket{U_+}}$}] (Up) at (-\latSize/2, {\latSize*(sqrt(3)/2)}){};
        \node[dot, label=above:{$\mathbf{\ket{V_+}}$}] (Vp) at (\latSize/2, {\latSize*(sqrt(3)/2)}){};
        \node[dot, label=60:{$\mathbf{\ket{T_+}}$}] (Tp) at (\latSize, 0){};
        \node[dot, label=120:{$\mathbf{\ket{T_-}}$}] (Tm) at (-\latSize, 0){};
        \node[dot, label=below:{$\mathbf{\ket{V_-}}$}] (Vm) at (-\latSize/2, -{\latSize*(sqrt(3)/2)}){};
        \node[dot, label=below:{$\mathbf{\ket{U_-}}$}] (Um) at (\latSize/2, -{\latSize*(sqrt(3)/2)}){};
        \fill (0,0) circle (6pt);
        \draw (0,0) circle (12pt);
        \node[yshift=3mm, xshift=7mm] at (0,0) {$\mathbf{\ket{T_3}}$};
        \node[yshift=-3mm, xshift=-6mm]  at (0,0) {$\mathbf{\ket{Y}}$};
        %LINES:
        \draw[dashed] (Up) -- (Tm) -- (Vm) -- (Um) -- (Tp) -- (Vp) -- (Up);
        \draw[dashed] (Um) -- (Up);
        \draw[dashed] (Vm) -- (Vp);
    \end{tikzpicture}
    \caption{Weight diagram of the adjoint representation.}
    \label{fig:adjointSU3Rep}
\end{figure}
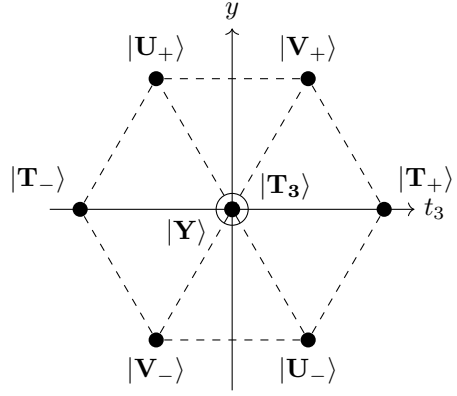

We can then put these states in a weight diagram (i.e. in a $(t_3, y)$ plane), which results in figure~\ref{fig:adjointSU3Rep}. Note how this parallels figure~\ref{fig:roots} above. This is of course no accident, since the states in the adjoint representation are the ladder (and diagonal $T_3$ and $Y$) operators themselves. One says that the roots are the weights of the adjoint representation.

Notice also that both $\ket{T_3}$ and $\ket{Y}$ have the same eigenvalues $(t_3=0,y=0)$. This does not mean that they are the same vector. Rather, it means that certain weights in an SU(3) representation may have multiplicity larger than one. In other words, there may be more than one state sitting on the same point of the weight diagram\footnote{One may wonder why this happens in SU(3) and not in SU(2) (since in that case two vectors with the same labels $j$ and $m$ are necessarily linearly dependent).
The reason for this difference stems from the fact that, in SU(2), there is essentially only one way to get from a state to another using the ladder operators: you can only "climb up or down" using $J_\pm$. In SU(3), on the other hand, you can reach one point from another by different paths in weight space. In other words, we can use several different combinations of $T_\pm, U_\pm$ and $V_\pm$ to reach one state from another, and these possibilities don't always commute, and will in general result in linearly independent states. Indeed, notice for instance that there are at least three possibilities to go from a state $(t_3, y)$ to $(t_3-1, y)$, namely by applying $T_-$, $U_+V_-$ or $V_-U_+$.
From the commutation relations in eq.~\eqref{eq:su3algebra} one sees that
\begin{equation*}
    [U_+,V_-] = T_- \implies U_+V_- = T_- + V_-U_+,
\end{equation*}
so $U_+V_-\ket{t_3,y}$, $T_-\ket{t_3,y}$ and $V_-U_+\ket{t_3,y} $ are linearly dependent, but any two of these are in general independent.}. In the case of the adjoint representation, the point $(0,0)$ has {multiplicity} 2. We represent this multiplicity in the weight diagram of figure~\ref{fig:adjointSU3Rep} by circling the central dot at the origin.

Because $T_-, T_3$ and $T_+$ form an $\mathfrak{su}(2)$ subalgebra of $\mathfrak{su}(3)$, the points along horizontal lines in SU(3) weight diagrams form SU(2) representations. Looking at figure~\ref{fig:adjointSU3Rep} we infer that: 
\begin{itemize}
    \item $\{\ket{U_+},\ket{V_+}\}$ form an SU(2) doublet,
   \item $\{\ket{T_-},\ket{T_3}, \ket{T_+}\}$ form an SU(3) triplet,
   \item $\{\ket{U_+},\ket{V_+}\}$ form an SU(2) doublet,
   \item $\ket{Y}$ forms an SU(2) singlet.
\end{itemize}
We know that the pions fit into an SU(2) triplet, so we could naturally try to accommodate them as the $\{\ket{T_-}, \ket{T_3}, \ket{T_+}\}$ in the octet. Moreover, recalling the discussion at the end of section~\ref{sec:zoo}, we know that $(K^0, K^+)$ and $(K^-, \overline{K^0})$ fit into SU(2) doublets, so we can associate them to $(\ket{U_+}, \ket{V_+})$ and $(\ket{V_-}, \ket{U_-})$. There remains the isospin singlet, which should correspond to another meson. In 1958, theorists who worked on the Sakata model predicted the existence of this new particle~\cite{Okun:1957, Yamaguchi:1958}, a new meson later dubbed $\eta$, which was indeed found with the properties anticipated from symmetry arguments~\cite{Bastien:1962zz}. In particular, note that its mass
\begin{equation*}
    m_\eta = 547.862 \pm 0.017~\text{MeV}
\end{equation*}
is very much in line with those of the kaons,
\begin{equation*}\begin{split}
    m_{K^\pm} = 493.677\pm 0.016~\text{MeV},\\
    m_{K^0} = m_{\overline{K}^0} = 497.611\pm 0.013~\text{MeV},
\end{split}\end{equation*}
in agreement with expectations from symmetry arguments (concerning their belonging to the same SU(3) multiplet).

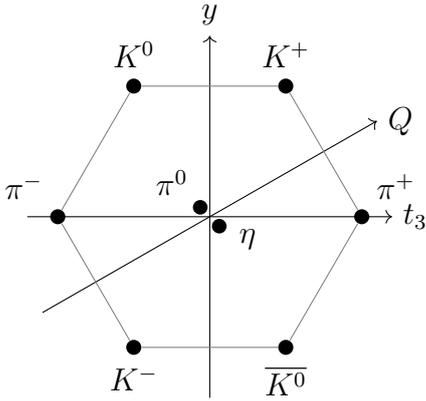
\begin{figure}
    \centering
    \begin{tikzpicture}[scale=0.5, dot/.style={circle,fill=black, inner sep=2pt}]
        \large
        \pgfmathsetmacro{\latSize}{4}

        \draw[->] (0, -1.2*\latSize) -- node[above, at end] {$y$} (0, 1.2*\latSize);
        \draw[->] (-1.2*\latSize, 0) -- node[right, at end] {$t_3$} (1.2*\latSize, 0);
        \draw[->] (-1.1*\latSize, -.635*\latSize) -- node[right, at end] {$Q$} (1.1*\latSize, .635*\latSize);

        \node[dot, label=above:{$K^0$}] (K0) at (-\latSize/2, {\latSize*(sqrt(3)/2)}){};
        \node[dot, label=above:{$K^+$}] (Kp) at (\latSize/2, {\latSize*(sqrt(3)/2)}){};
        \node[dot, label=60:{$\pi^+$}] (pip) at (\latSize, 0){};
        \node[dot, label=120:{$\pi^-$}] (pim) at (-\latSize, 0){};
        \node[dot, label=below:{$K^-$}] (Km) at (-\latSize/2, -{\latSize*(sqrt(3)/2)}){};
        \node[dot, label=below:{$\overline{K^0}$}] (K0bar) at (\latSize/2, -{\latSize*(sqrt(3)/2)}){};
        \fill (-.25,.25) circle (5.5pt);
        \fill (.25,-.25) circle (5.5pt);
        \node[above=4pt, xshift=-5mm] at (0,0) {$\pi^0$};
        \node[yshift=-3mm, xshift=5mm]  at (0,0) {$\eta$};
        %LINES:
        \draw[color=gray] (K0) -- (Kp) -- (pip) -- (K0bar) -- (Km) -- (pim) -- (K0);
\end{tikzpicture}
    \caption{The meson octet. Note how the charge increases in the direction of the $Q$ axis. As we move in the direction perpendicular to this axis the charge remains unaltered.}
    \label{fig:meson8}
\end{figure}

Figure~\ref{fig:meson8} illustrates this fitting of the mesons in the octet representation. Note that the electric charge of the particles do not change as we move along the direction of the $U_\pm$ roots (or as we apply the $U_\pm$ ladders). Instead, charge increases or decreases as we move in the orthogonal direction, which is also shown in the figure as the $Q$ axis, corresponding to the operator\footnote{Note that this is perpendicular to the direction $U_3\sim\lambda_3 + \sqrt{3}\lambda_8$, which is another Pauli matrix that can be built out of the Gell-Mann matrices. We chose to work with $T_3$ and $Y$ as the diagonal generators, but we could with equal right have chosen $U_3$ and $Q$. This is yet another illustration of the fact that $\mathfrak{su}(3)$ is formed of 3 overlapping $\mathfrak{su}(2)$'s. Note also that the states lying on top of the $U_3$ direction belong to $\mathfrak{su}(2)$ representations, which are symmetric around the origin. Thus the weight diagram must be symmetric against reflections around $Q$, as can be indeed seen from figure~\ref{fig:meson8}.}
\begin{equation}
    Q \equiv T_3 + \frac{Y}{2}.
    \label{eq:QY}
\end{equation}
Comparing with eq.~\eqref{eq:GMNishijima} we see that $Y=B+S$, and since for mesons $B=0$ we see that in this case the hypercharge $Y$ is just a quantification of strangeness. This is again in line with our discussions of section~\ref{sec:zoo}, where we concluded that $(K^0, K^+)$ doublet had strangeness $S=+1$ while their antiparticles $(\overline{K^0}, K^-)$ have $S=-1$. 

\subsubsection*{The decuplet}

\begin{figure}
    \centering
    \begin{tikzpicture}[scale=0.4, dot/.style={circle,fill=black, inner sep=2pt}]
        \large
        \pgfmathsetmacro{\latSize}{4}

        \draw[->] (0, -2.2*\latSize) -- node[above, at end] {$y$} (0, 1.2*\latSize);
        \draw[->] (-1.2*\latSize, 0) -- node[right, at end] {$t_3$} (1.2*\latSize, 0);
        \draw[->] (-1*\latSize, {-\latSize*sqrt(3)/3}) -- node[right, at end] {$Q$} (1.8*\latSize, {\latSize*.6*sqrt(3)});

        \node[dot, label=above:{$\Delta^{++}$}] (dpp) at (1.5*\latSize, {\latSize*(sqrt(3)/2)}){};
        \node[dot, label=above:{$\Delta^+$}] (dp) at (.5*\latSize, {\latSize*(sqrt(3)/2)}){};
        \node[dot, label=above:{$\Delta^0$}] (d0) at (-.5*\latSize, {\latSize*(sqrt(3)/2)}){};
        \node[dot, label=above:{$\Delta^-$}] (dm) at (-1.5*\latSize, {\latSize*(sqrt(3)/2)}){};
        \node[dot, label=150:$\Sigma^{*-}$] (sm) at (-\latSize, 0){};
        \node[dot, label=150:$\Sigma^{*0}$] (s0) at (0,0){};
        \node[dot, label=30:$\Sigma^{*+}$] (sp) at (\latSize, 0){};
        \node[dot, label=left:$\Xi^{*-}$] (xim) at (-0.5*\latSize, -{\latSize*(sqrt(3)/2)}){};
        \node[dot, label=right:$\Xi^{*0}$] (xi0) at (0.5*\latSize, -{\latSize*(sqrt(3)/2)}){};
        \node[dot, label=right:$\Omega^-$] (Om) at (0, -{\latSize*(sqrt(3))}){};
        %LINES:
        \draw[color=gray] (dpp) -- (sp) -- (xi0) -- (Om) -- (xim) -- (sm) -- (dm) -- (d0) -- (dp) -- (dpp);
\end{tikzpicture}
    \caption{The decuplet fits the spin $3/2$ baryons nicely. When this classification scheme was first proposed the $\Omega^-$ baryon was still not known. However, symmetry arguments allowed theorists such as Gell-Mann, Glashow and Sakurai to predict that this particle should exist, with spin $3/2$, charge $-1$ and mass approximately equal to that of the other baryons of the decuplet. A few years later the particle was indeed discovered, which is accounted as one of the greatest triumphs of this classification scheme based on SU(3).}
    \label{fig:decuplet}
\end{figure}
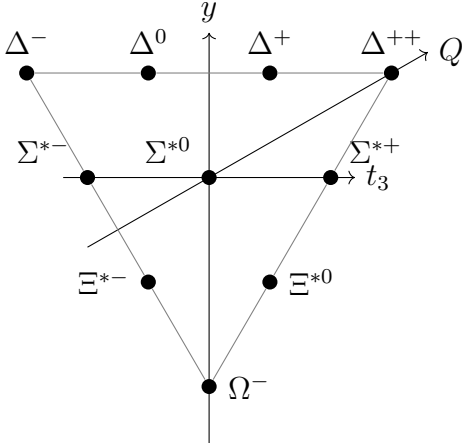

Finally, another important SU(3) representation is the decuplet, thus named because it is a 10-dimensional vector space. Its weight diagram is shown in figure~\ref{fig:decuplet}. It can be constructed by starting at a state with largest value of $t_3$ (this state must exist, otherwise the representation space would be infinite dimensional) and applying $V_-$ transformations 3 times consecutively. By the fourth application of $V_-$ the result is the null vector. This is akin to the case of SU(2), when a state with $m=m_\text{min}$ had to be annihilated by the lowering operator for consistency. Moreover, the decuplet is such that, after applying three times $V_-$, one cannot do other operations except go back up via $V_+$ or via $U_+$. It can be shown that $U_+$ can be applied only 3 times before annihilating the state, and then one can only apply $T_+$ to close the triangle, as shown in figure~\ref{fig:decuplet}~\cite{Gasiorowicz:1966xra}.

The decuplet is important in particle physics because, up until 1963, we had detected nine spin $3/2$ baryons, namely: the four $\Delta$'s that we met in section~\ref{sec:quantitative}, a triplet of baryons called $\Sigma^*$, and a doublet of $\Xi^*$~\cite{Barnes:1964pd}. Their properties fit nicely into the decuplet representation. 

From eq.~\eqref{eq:QY} we see, for example, that the state at the top-right corner of the triangle has $Y=1$ and $T_3=3/2$, resulting in $Q=2$. We did know a particle with this charge: the $\Delta^{++}$ baryon, with isospin $3/2$ under SU(2). As discussed in section~\ref{sec:quantitative} we expect four such baryons, which were indeed found and fit nicely into the top-most row of the triangle in figure~\ref{fig:decuplet}. 

As we move to the left in the diagram along a horizontal line, the charge $Q$ decreases by one unit. Likewise, descending in the $V_-$ direction decreases $Y$ by $1$ and $T_3$ by $1/2$, therefore decreasing $Q$ by unit as well. So the $\Sigma^*$'s fit naturally in the row below the $\Delta$'s, and the $\Xi^*$'s fit in the row below the $\Sigma^*$'s.

Moreover, recalling that $Y=B+S$ and that these particles are baryons, one has $S=Y-1$ in this case. This prediction is also in agreement with the observations that $S_\Delta=0$, $S_{\Sigma^*}=-1$ and $S_{\Xi^*}=-2$.

However, symmetry arguments tell us that we should have 10 of these particles, and not just nine. This led many theorists in 1962 to conjecture that a tenth particle should exist~\cite{GellMann:1962, Glashow:1962zza}, dubbed the $\Omega$ baryon (since $\Omega$ is the last letter of the greek alphabet, and this would be the last missing member of the decuplet). Moreover, the properties of this particle can be predicted from symmetry arguments: it should have spin $3/2$, charge $-1$ and strangeness $-3$. Not only that, the mass splitting between the consecutive rows of the diagram is approximately constant, since
\begin{align*}
    m_\Delta &\simeq 1232~\text{MeV},\\
    m_{\Sigma^*}&\simeq 1385~\text{MeV},\\
    m_{\Xi^*} &\simeq 1532~\text{MeV},
\end{align*}
allowing us to also predict
\[
    m_{\Omega^-}^\text{pred.} \simeq 1680~\text{MeV}.
\]
In February 1964 the discovery of a particle with exactly these properties was announced~\cite{Barnes:1964pd}, with mass $m_{\Omega^-}^\text{exp.} = 1672.45$~MeV (within $\sim 0.5\%$ of the predicted mass!). To this day the prediction and subsequent detection of the $\Omega^-$ baryon is one of the greatest success stories in particle physics.

\subsection{Tensor products of SU(3) representations}
\label{sec:tensorsu3}

We can also consider the analogous problem to what we studied in section~\ref{sec:sum}: what happens when one deals with a composite system of two (or more) particles belonging to SU(3) representations? The resulting composite system must also belong to SU(3) representations. But which ones? Mathematically, this amounts to determining how tensor products of SU(3) representations decompose as direct sums of SU(3) representations.

For this purpose, note that the developments that led to eqs.~\eqref{eq:Ux} and \eqref{eq:sumJi} are valid for any group. So also in SU(3) the eigenvalues $t_3$ and $y$ of a composite system $\ket{t_3^{(1)}, y^{(1)}}\otimes\ket{t_3^{(2)}, y^{(2)}}$ are $t_3^{(1)} + t_3^{(2)}$ and $y^{(1)} + y^{(2)}$.

\begin{figure}
    \centering
    \begin{equation*}
        \mathbf{3\otimes\overline{3}} = 
        \vcenter{\hbox{
            \begin{tikzpicture}[scale=0.3, dot/.style={circle,fill=black, inner sep=1pt}]
                \pgfmathsetmacro{\latSize}{4}
                % DRAW THE AXES
                \draw[->] (0, -\latSize) -- node[above, at end] {$y$} (0, \latSize);
                \draw[->] (-\latSize, 0) -- node[right, at end] {$t_3$} (\latSize, 0);

                \node[dot, color=red] (u) at ({\latSize*(1/2)}, {\latSize*(sqrt(3)/6)}){};
                \node[dot, color=red] (d) at ({\latSize*(-1/2)}, {\latSize*(sqrt(3)/6)}){};
                \node[dot, color=red] (s) at (0, {-\latSize*(sqrt(3)/3)}){};
                % DEFINE THE COORDINATES OF THE POINTS IN THE ADJOINT REP (up, down, strange and their anti-particles)
                \node[dot] (das) at (-\latSize/2, {\latSize*(sqrt(3)/2)}){};
                \node[dot] (dau) at (-\latSize, 0){};
                \node[dot] (sau) at (-\latSize/2, -{\latSize*(sqrt(3)/2)}){};
                \node[dot] (sad) at (\latSize/2, -{\latSize*(sqrt(3)/2)}){};
                \node[dot] (uad) at (\latSize, 0){};
                \node[dot] (uas) at (\latSize/2, {\latSize*(sqrt(3)/2)}){};
            
                % DRAW THE ANTI-FUNDAMENTAL REP AT up
                \fill[black]    (.25,.25) circle (5pt); 
                \draw[black][-] (0,0) -- (uad) -- (uas) -- (0,0);
                
                % DRAW THE ANTI-FUNDAMENTAL REP AT down
                \fill[black]    (-.25,.25) circle (5pt);
                \draw[black][-] (0,0) -- (das) -- (dau) -- (0,0);

                % DRAW THE ANTI-FUNDAMENTAL REP AT strange
                \fill[black]    (0,-.25) circle (5pt);
                \draw[black][-] (0,0) -- (sau) -- (sad) -- (0,0);

                % DRAW DOTTED LINES CONNECTING THE POINTS OF THE FUNDAMENTAL
                \draw[dotted, color=red] (s) -- (d) -- (u) -- (s);                
            \end{tikzpicture}
        }}
    \end{equation*}
    \begin{equation*}
        = 
        \vcenter{\hbox{
            \begin{tikzpicture}[scale=0.3, dot/.style={circle,fill=black, inner sep=1pt}]
                \pgfmathsetmacro{\latSize}{4}
                \node[dot] (das) at (-\latSize/2, {\latSize*(sqrt(3)/2)}){};
                \node[dot] (dau) at (-\latSize, 0){};
                \node[dot] (sau) at (-\latSize/2, -{\latSize*(sqrt(3)/2)}){};
                \node[dot] (sad) at (\latSize/2, -{\latSize*(sqrt(3)/2)}){};
                \node[dot] (uad) at (\latSize, 0){};
                \node[dot] (uas) at (\latSize/2, {\latSize*(sqrt(3)/2)}){};
                \fill (0,0) circle (5pt);
                \draw (0,0) circle (10pt);                
                \draw[-] (dau) -- (das) -- (uas) -- (uad) -- (sad) -- (sau) -- (dau);
            \end{tikzpicture}
        }}
        \oplus
        \vcenter{\hbox{
            \begin{tikzpicture}[scale=0.3]
                \pgfmathsetmacro{\latSize}{4}
                % DRAW THE AXES
                \draw[->] (0, -\latSize) -- node[above, at end] {$y$} (0, \latSize);
                \draw[->] (-\latSize, 0) -- node[right, at end] {$t_3$} (\latSize, 0);
                % DRAW THE TRIVIAL REP
                \fill (0,0) circle (8pt);
            \end{tikzpicture}
        }}
    \end{equation*}
    \caption{Diagramatic illustration of tensor product of the fundamental and the antifundamental representation, resulting in the octet and the singlet.}
    \label{fig:33bar}
\end{figure}
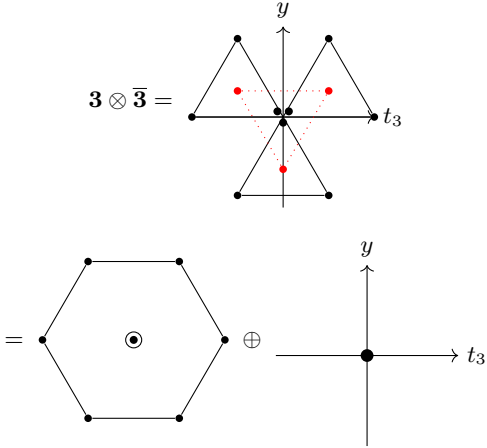

In the $(t_3,y)$ weight diagram, this means that when we take the tensor product of two representations, the resulting figure is achieved by superimposing copies of one of these representations over each point belonging to the other representation (i.e. taking the weights of one representation as the origin of the other's weight diagram). Figure \ref{fig:33bar} illustrates this for the tensor product of the fundamental and the antifundamental representations. We see that the outer points form a hexagon, and there are three weights at the origin. The octuplet is an hexagon with multiplicity two at the origin, so we see that the 9-dimensional space obtained from this tensor product naturally decomposes as
\begin{equation*}
    \mathbf{3}\otimes \mathbf{\overline{3}} = \mathbf{8}\oplus \mathbf{1}.
\end{equation*}

With the same construction we can perform the product of $\mathbf{3}\otimes \mathbf{3}$ and identify a $\mathbf{\overline{3}}$ as well as a triangle corresponding to a 6-dimensional representation which we call the $\mathbf{6}$. Likewise we can see that $\mathbf{3}\otimes \mathbf{6}$ results in an octet and a decuplet~\cite{Thyssen:2017}. So we can write
\begin{equation*}\begin{split}
    \mathbf{3\otimes 3\otimes 3} &= (\mathbf{6\oplus \overline{3}})\otimes \mathbf{3}\\
    &=(\mathbf{6}\otimes \mathbf{3})\oplus (\mathbf{\overline{3}\otimes 3})\\
    &= \mathbf{10\oplus 8\oplus 8 \oplus 1}.
\end{split}\end{equation*}

\subsection{Sakatons \emph{vs} quarks: triumphs of the quark model}

These results indicate that all SU(3) representations can be recovered from tensor products of the fundamental and antifundamental. It is precisely because they act as building blocks of all representations that they are called ``fundamental''. 

From figure~\ref{fig:33bar} we see that, in Sakata's model, the octet + singlet could then be seen as composite states of protons/neutrons/$\Lambda$'s with their antiparticles $\overline{p}$/$\overline{n}$/$\overline{\Lambda}$. These fundamental constituents of the Sakata model are often denoted as \emph{sakatons}. This perspective fits nicely with the meson octet in figure~\ref{fig:meson8}, since all quantum numbers would match if we identify $\pi^+ = \overline{n}p$, $K^+ = p\overline{\Lambda}$, $K^0=n\overline{\Lambda}$, $\pi^0=(\overline{n}n + \overline{p}p)/\sqrt{2}$ (the $m=0$ state of the SU(2) triplet), and $\eta = (\overline{n}n - \overline{p}p - 2\overline{\Lambda}\Lambda)/\sqrt{6}$. The remaining mesons of the octed are antiparticles of these. Note, for instance, how in this scheme the $S=+1$ kaons $K^0$ and $K^+$ would be composed of a $\overline{\Lambda}$, which is the antiparticle of an $S=-1$ particle.

One sees that the mesons fit in octets which we can interpret as stemming from $\mathbf{3\otimes \overline{3}}$. This product also contain a singlet (now called $\eta^\prime$), predicted since 1957~\cite{Okun:1957, Ikeda:1959zz} and observed in 1964~\cite{Kalbfleisch:1964zz, Goldberg:1964zza, Dauber:1964zz} -- yet another success for this mesonic classification scheme.

However, this model is a lot less successful for a classification of baryons. When Sakata proposed this model in 1956, eight spin-1/2 baryons were known: $p$, $n$, $\Lambda$, 3 $\Sigma$'s and 2 $\Xi$'s\footnote{Note that these are different particles from the spin-3/2 $\Sigma^*$'s and $\Xi^*$'s appearing in the decuplet. The similar names come from similarities they share, e.g. the fact that both $\Sigma$ and $\Sigma^*$ exist as isospin triplets under SU(2), i.e. there are three isospin partners ($\Sigma^-, \Sigma^0, \Sigma^+)$ and also three isospin partners ($\Sigma^{*-}, \Sigma^{*0}, \Sigma^{*+})$. Similarly, both $\Xi$ and $\Xi^*$ form SU(2) doublets.}~\cite{Maki:1961, Okun:2006nq}. Since the $\Sigma$'s have strangeness $S_{\Sigma}=-1$, they would have to contain one $\Lambda$. Clearly a $\Sigma^-$ would have to also include an antiproton $\overline{p}$ (since this is the only sakaton with negative charge), so we would need at least a $\mathbf{3\otimes \overline{3}} = \mathbf{8}\oplus \mathbf{1}$. But there is no way to fit the $\Sigma$ triplet in the octet: the only $t_3$-triplet in this representation lies on the $t_3$ axis, and not all states in this axis would contain a $\Lambda$ (meaning not all states would have the correct strangeness for the $\Sigma$'s), see fig.~\ref{fig:33bar}. We could then try to accommodate the $\Sigma$'s and $\Xi$'s using larger representations, for instance $\mathbf{3\otimes 3\otimes \overline{3}}$, which is $3\times 3\times 3 = 27$-dimensional. However, the other 22 baryonic partners of the $\Sigma$'s and $\Xi$'s, which would be expected to belong to this large representation space, were never found. 

\begin{figure}
    \centering
    \begin{tikzpicture}[scale=0.5, dot/.style={circle,fill=black, inner sep=2pt}]
        \large
        \pgfmathsetmacro{\latSize}{4}

        \draw[->] (0, -1.2*\latSize) -- node[above, at end] {$y$} (0, 1.2*\latSize);
        \draw[->] (-1.2*\latSize, 0) -- node[right, at end] {$t_3$} (1.2*\latSize, 0);
        \draw[->] (-1.1*\latSize, -.635*\latSize) -- node[right, at end] {$Q$} (1.1*\latSize, .635*\latSize);

        \node[dot, label=above:{$n$}] (K0) at (-\latSize/2, {\latSize*(sqrt(3)/2)}){};
        \node[dot, label=above:{$p$}] (Kp) at (\latSize/2, {\latSize*(sqrt(3)/2)}){};
        \node[dot, label=60:{$\Sigma^+$}] (pip) at (\latSize, 0){};
        \node[dot, label=120:{$\Sigma^-$}] (pim) at (-\latSize, 0){};
        \node[dot, label=below:{$\Xi^-$}] (Km) at (-\latSize/2, -{\latSize*(sqrt(3)/2)}){};
        \node[dot, label=below:{$\Xi^0$}] (K0bar) at (\latSize/2, -{\latSize*(sqrt(3)/2)}){};
        \fill (-.25,.25) circle (5.5pt);
        \fill (.25,-.25) circle (5.5pt);
        \node[above=4pt, xshift=-5mm] at (0,0) {$\Sigma^0$};
        \node[yshift=-3mm, xshift=5mm]  at (0,0) {$\Lambda$};
        %LINES:
        \draw[color=gray] (K0) -- (Kp) -- (pip) -- (K0bar) -- (Km) -- (pim) -- (K0);
\end{tikzpicture}
    \caption{The spin-1/2 baryon octet of the ``Eightfold Way''.}
    \label{fig:baryon8}
\end{figure}
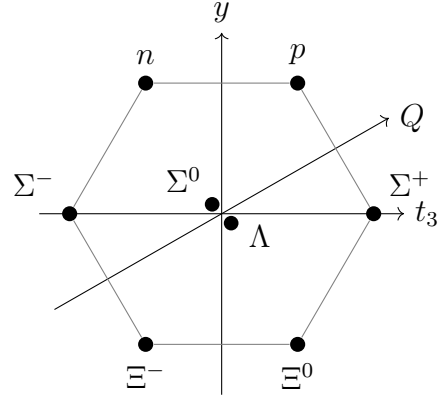

It was Gell-Mann \cite{Gell-Mann:1961TheEightFold} and Ne'emann \cite{Neeman:1961TheEightFold}, working independently, who first proposed a different baryonic classification scheme. Instead of taking $p, n$ and $\Lambda$ as belonging to the fundamental, they instead proposed that these should be classified together with the $\Sigma$'s and $\Xi$'s in the $\mathbf{8}$ representation. This is the famous ``Eightfold Way'' of particle physics, depicted in fig.~\ref{fig:baryon8}. 

In this scenario, the spin-1/2 baryons fit in an octet while the spin-3/2 ones fit the decuplet. Looking at the developments of the last section, we can identify this $\mathbf{10\oplus 8}$ as a result\footnote{The reader may now look at the expression written at the end of section~\ref{sec:tensorsu3}, namely $\mathbf{3\otimes 3\otimes 3 = 10\oplus 8 \oplus 8 \oplus 1}$, and conclude that we should find one baryonic decuplet, two baryonic octets and one baryonic singlet. But only one decuplet and one octet are actually found! A detailed explanation of why the $\mathbf{8\oplus 1}$ baryons do not exist is beyond the scope of this paper. It is related to the confinement mechanism of quarks. It turns out that quarks carry not only flavour (i.e. their characteristic of being an $u$, $d$ or $s$) but also spin 1/2 and a colour charge. Quark confinement means that only fully anti-symmetrized wavefunctions in the colour degree of freedom are allowed. By Pauli's exclusion principle this means that the wavefunction must be fully symmetric in flavour and spin. It turns out that only the $\mathbf{10\oplus 8}$ satisfy these requirements, and are therefore the only SU(3) baryonic representations that can be found in Nature~\cite{Halzen:1986}.} of the product $\mathbf{3\otimes 3\otimes 3}$.

The remaining question is: what is the fundamental representation in this case? Which are the fundamental constituent particles of mesons and baryons? Gell-Mann dubbed these particles ``quarks''\cite{Gell-Mann:1964Quarks}\footnote{George Zweig also had a similar idea, and dubbed these fundamental constituents ``aces''~\cite{Zweig:1964}. Clearly it was Gell-Mann's nomenclature that stuck. Incidentally, Gell-Mann took the idea of the name from a passage of James Joyce's \emph{Finnegans Wake}: ``Three quarks for Muster Mark!'', inspired by the $\mathbf{3\otimes 3\otimes 3}$ construction of baryons.}. The three states of the fundamental representation are called \emph{up}, \emph{down} and \emph{strange} quarks, or, in terms of $t_3$ and $y$ eigenvalues,
\[
    u = \left|\frac12, \frac13\right\rangle,\quad 
    d = \left|-\frac12, \frac13\right\rangle,\quad
    s = \left|0, -\frac23\right\rangle.
\] 
A few important conclusions follow straightforwardly, namely:
\begin{itemize}
    \item applying eq.~\eqref{eq:QY} to the fundamental representation one sees that these quarks have charges $Q_u = 2/3$, $Q_d=-1/3$ and $Q_s = -1/3$. Their charges are thus fractions of the elementary charge.
    \item Since baryons are constituted of 3 quarks, these constituents must have fractional baryonic number $B=1/3$. Mesons are formed of quarks and anti-quarks and thus have $B=1/3 + (-1/3) = 0$.
    \item Looking at the meson octet in fig.~\ref{fig:meson8} and how it stems from the product $\mathbf{3\otimes \overline{3}}$ as in fig.~\ref{fig:33bar}, one concludes that the strangeness charge of the mesons is nothing but a count of the number of strange quarks (or rather strange antiquarks, since $s$ has strangeness $-1$) that constitute them. Likewise for the baryons in the octet and in the decuplet.
    \item The fact that strangeness is only approximately conserved can now be interpreted in the following manner: the strong interaction is not able to change the nature of a quark, but the weak interaction is. So, in a strangeness-violating decay such as $\Lambda^0\to p+\pi^-$, what happens is that the strange quark in $\Lambda^0 = uds$ is decaying as 
    \begin{align*} 
        s\to u + 
        &
        \underset{|}{W^-}\\[-4mm]
        &~\,\rightarrow d+\overline{u}
    \end{align*}
    with $W^-$ a mediator of the weak interaction. Then
    \[
        \Lambda^0 = uds \to {uud} + {d\overline{u}} = p + \pi^-.
    \] 
\end{itemize}

Despite these predictive successes, for some time there was some skepticism as to whether quarks indeed exist, or whether they would just be mathematical artifacts of this symmetry construction~\cite{Gasiorowicz:1966xra}. For one, it is relatively easy to detect charged particles, but despite decades of successful experiments in particle physics no particles with fractional charges had ever been seen. It was then clear that quarks, if they existed, could never be seen in isolation, but only confined within hadrons -- and this confinement mechanism was largely unknown at the time (and still is to a great extent!). 

By 1968, high-energy collisions of electrons and nuclei were finally able to probe the inner structure of hadrons, showing that they indeed behave as if constituted of point-like substructures. Throughout the early 1970s a series of other theoretical and experimental discoveries helped explain some aspects of quark confinement, and strengthened the conviction of the community on the existence of these fundamental hadronic constituents~\cite{Riordan:1992hr}. Overall, this ultimate triumph of the quark model was another chapter in the long history of theoretical prediction of new particles followed by subsequent experimental confirmation.

\section{Conclusions}
\label{sec:conclusions}

Symmetry arguments and related techniques from group representation theory have always played a pivotal role in the development of particle physics. These techniques appear in the study of the Lorentz group, in classification schemes proposed since the earliest days of particle physics, and in the \emph{gauge principle} that now lies at the heart of the Standard Model of elementary particles and their fundamental interactions. It is therefore essential that any beginner in this field be familiar with this language.

In this paper we paved a pedagogic path that serves as an introduction to group representation theory for beginner particle physicists. Our approach focuses in the history of hadronic classification schemes that culminated in the quark model in the mid- to late 1960s. We argued that this is the most natural starting point to introduce group theory to this audience, since it allows one to start from the most basic SU(2) group and progress naturally to the more challenging SU(3) structure. Moreover, this allows students to get acquainted with some aspects of the early history of this field and some of its most successful predictions. Through the study of symmetries and groups we can tell the story of many particles that were predicted and later detected, a constant pattern that is still ongoing in the field of particle physics.

This approach is also particularly interesting because it allows beginners to have a first contact with the history of the quark model. Most students are already familiar with the fact that protons and neutrons are composed of quarks. But as scientists it is important to not only know something as a fact, but also to know the details of \emph{how and why we came to know that as a fact}. And for the quark model this inevitably means discussing SU(3) representations and/or deep inelastic scattering experiments.

In this sense, this work differs from most of the current literature that choose to present the SU(3) paradigm already from an \emph{ad hoc} postulate of the existence of 3 quark flavours. Here we instead culminate in the \emph{prediction} of quarks as belonging to the fundamental representation of a (previously recognized) SU(3) symmetry group. In this pedagogic path, quarks are yet another paradigmatic example of particles predicted from symmetry arguments and later confirmed by experiments.

Finally, an important novelty in this presentation is a middle path we have found between a full presentation of mathematical tools and an emphasis on physical applications. The reader will find here an explanation of the concept of group representation, including the fundamental and the adjoint, as well as the main techniques underlying the highest weight construction which can be used to build general representations of Lie algebras. This allows for a reasonable understanding of the ideas being discussed, and provides the reader with tools to later explore more technical texts. On the other hand, we purposefully avoid any general results, sticking to the groups we are actually interested in. Abstract mathematical definitions of groups and representations are reserved for the appendix. Theorems on the geometry of SU(3) weight diagrams are not stated nor proven. We have focused only on the results that are necessary for the reader to understand the ``Eightfold Way'' and the quark model.

After studying the subject as presented here, the reader will hopefully be more prepared to tackle more thorough textbooks on group theory, with the benefit of knowing where some of these arid concepts will appear and how they can be applied. Should this text help in providing students with the toolkit to face this journey more lightly, it will have fulfilled its purpose.

\section*{Acknowledgements}

G.F.V would like to thank FAPEMIG and Universidade Federal de Minas Gerais (UFMG) for financial support during the preparation of this work.

\appendix

\section{Formal definition of representations}

In the presentation above we have mentioned the concept of ``groups'', but have purposefully avoided defining it rigorously, since one can understand the underlying physical concepts without going into all the mathematical details. The aim of this appendix is to partially fill this gap to the interested reader.

\subsubsection*{Definition of group}

As discussed above, we are interested in studying how certain transformations act on physical systems. These transformations are such that we can perform a number of them one after the other, and the net outcome is still a transformation of this same type. Mathematically we say that, in the set of all such transformations, we can define a binary operation that takes two transformations and composes them into a third. More formally, the transformations form a set $G$ where we can define an operation $G\times G\to G$. This composition is such that: (i) there is an ``identity'' transformation $e\in G$, which is the operation of ``do nothing to the system'', such that if we compose any transformation $g$ with the identity $e$ the result is $g$ itself (i.e. if we do one transformation $g$, and then do nothing else, the overall transformation is just $g$ itself); (ii) after we perform a transformation $g$, we can always perform an inverse transformation $g^{-1}$ that takes us back to the initial state, i.e. such that composing $g$ with $g^{-1}$ is the ``do nothing'' transformation; (iii) the composition is associative, i.e. composing $g_1$ with $g_2$ and then composing this outcome with $g_3$ is the same as composing $g_1$ with the outcome of $g_2$ and $g_3$. Note that, in principle, we only know how to compose \emph{pairs} of transformations. But property (iii) tells us that we can compose multiple transformations by separating them in pairs, and the outcome doesn't depend on how we do this splitting. Any set with a binary operation that satisfies these properties is called a \emph{group}.

\subsubsection*{Group representations}

We are not so much interested in the transformations themselves, but rather on how they act on physical systems. These are typically described by vectors in a linear space (such as the Hilbert space of the system in quantum mechanics). Therefore we want to understand how to describe these transformations as matrices (i.e. linear operators) acting on vector spaces. In other words, we want a map 
\[ D: G\to \text{Aut}(V)\]
of the abstract transformation group $G$ to the set of invertible\footnote{They must be invertible because every transformation has an inverse, as per the definition of groups.} operators acting on the state space $V$ of the system. Not only that, but this map $D$ must preserve the multiplication structure of the group, i.e.
\begin{equation}
    D(g_1) D(g_2)=D(g_1\cdot g_2).
    \label{eq:homomorphism}
\end{equation} 
What this tells us is that the matrix multiplication of $D(g_1)$ and $D(g_2)$ coincides with the matrix representing the composite transformation $g_1\cdot g_2$. We say that $D$ preserves the group multiplication structure.  
Any map $D:G\to \text{Aut}(V)$ preserving this group multiplication is called e \emph{representation} of the group acting on the space $V$.  
A good analogy is to understand this concept of representation as photographs or maps of the group in question: a photo or a map illustrates the place to which they refer, its buildings and geographical structures, but are not the place itself~\cite{Zee:2016fuk}.

How is this definition related to our use of the word \emph{representation} in the main text? To find a representation of the group is to find a map $D:G\to \text{Aut}(V)$ as defined above, for some vector space $V$. But this then means that all transformations map $V$ into itself, i.e. the space $V$ contain all possible states that can be obtained one from the other via such transformations. This is exactly what we have done in sections~\ref{sec:repssu2} and \ref{sec:repsSU3}: we have constructed spaces $V$ of states that are related among themselves by transformations.

\subsubsection*{(Ir)reducibility}

It may happen that the space $V$ is actually too large, in the sense that it has subspaces that remain invariant under the transformations in question. Suppose $W\subset V$ is such an invariant subspace. This means that the elemenets of $W$ transform among themselves, and there is no transformation that can take an element of $W$ outside of $W$. Think, for example, of the total state space of nucleons and pions. Isospin transformations can take a $\pi^-$ to $\pi^+$, or a $\pi^0$ to a $\pi^-$, but there is no way to transform a pions into a nucleon by an isospin transformation: they actually belong to different (totally unrelated) representation spaces. So a space of states of pions and nucleons can actually be \emph{reduced} to two separate spaces: that of the nucleon isospin states, and that of the pion states.

When this happens, it means we can write this reducible state space as a direct sum of the independent pieces, i.e. $V_1\oplus V_2$. If $D^1$ is a representation on $V_1$ and $D^2$ acts on $V_2$, then
\begin{equation*}
    D^{V_1\oplus V_2}(g)\begin{pmatrix}
        \ket{v_1}\\
        \ket{v_2}
    \end{pmatrix} \equiv 
    \begin{pmatrix}
        D^1(g)(\ket{v_1})\\
        D^2(g)(\ket{v_2})
    \end{pmatrix}.
\end{equation*}
This means that in matrix form $D^{V_1\oplus V_2}$ takes a block diagonal shape
\begin{equation*}
    D^{V_1\oplus V_2}(g) =
    \left(\begin{array}{ccccc}
	  \multicolumn{2}{c|}{\multirow{3}{*}{$D^1(g)$}} &\multicolumn{3}{c}{\multirow{3}{*}{$\mathbf{0}$}} \\ 
        \multicolumn{2}{c|}{}                          &\multicolumn{3}{c}{} \\
        \multicolumn{2}{c|}{}                         & \multicolumn{3}{c}{} \\ \cline{1-5}
        \multicolumn{2}{c}{} & \multicolumn{3}{|c}{} \\
        \multicolumn{2}{c}{} & \multicolumn{3}{|c}{}  \\
        \multicolumn{2}{c}{\raisebox{5pt}{~\quad $\mathbf{0}$\quad~~}} & \multicolumn{3}{|c}{\raisebox{5pt}{~\quad $D^2(g)$\quad~~}} \\
        \multicolumn{2}{c}{} & \multicolumn{3}{|c}{}  \\ 
	\end{array}\right).
\end{equation*}

This is exactly what happened in eq.~\eqref{eq:U}. We can treat $V_1$ and $V_2$ as separate spaces, and ask again if the same reducibility can be performed, i.e. if they have invariant subspaces other than themselves. If not, then the representation on $V_1$ (resp. $V_2$) is said to be \emph{irreducible}. 

When a representation is irreducible, it means that every state is related to any other via some transformation. This is exactly what we have constructed in sections~\ref{sec:repssu2} and \ref{sec:repsSU3}: we have thus built \emph{irreducible} representations of SU(2) and SU(3).

\subsubsection*{Tensor product representation}

Given two vector spaces $V$ and $W$, with basis $\{\ket{v_i}\}$ and $\{\ket{w_j}\}$ respectively, we can build another space with a basis $\{\ket{v_i}\otimes \ket{w_j}$, where $\otimes$ can be seen as just a symbol for composing states of $V$ and $W$ into a single vector. The resulting $\text{dim}(V)\times \text{dim}(W)$-dimensional space is denoted the \emph{tensor product} $V\otimes W$.

If $D^V$ (resp. $D^W$) is a representation over $V$ (resp. $W$), it is not difficult to show that the map $D^{V\otimes W}$ defined by
\begin{equation*}
    D^{V\otimes W}(g)(\ket{v_i}\otimes \ket{w_j}) \equiv D^V(g)(\ket{v_i}) \otimes D^W(g)(\ket{w_j})
\end{equation*}
is also a representation (i.e. it satisfies property~\eqref{eq:homomorphism} above). This is called the \emph{tensor product representation}. It just means that a composite state belonging to a tensor product transforms in the following way: the transformed state from $\ket{v}\otimes \ket{w}$ is obtained by separately transforming the states in their natural subspaces $V$ and $W$, and then taking their tensor product. This is exactly what we did in eq.~\eqref{eq:Ux} above.

\section{From the algebra to the group}

In section~\ref{sec:su2inftl} we have explicitly said that, because we are dealing with a set of continuous groups (both in the case of SU(2) isospin as well as in extended isospin of SU(3)), we could construct representations by investigating how states transform under on infinitesimal transformations -- which is what we did throughout the rest of the paper. Let us briefly discuss this in slightly more formal terms.

\begin{figure}
    \centering
    \includegraphics[scale=.4]{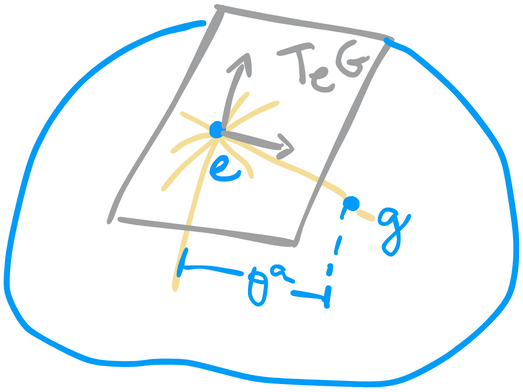}
    \caption{Illustration of a Lie group, a group of continuous transformations. The element $g$ is connected to the identity via a continuous path of points (every point on this path corresponds to some transformation that can be performed over the system). The tangent vector along this curve is $T_a$ and the coordinate of this point $g$ is $\theta^a$.}
    \label{fig:Lie}
\end{figure}

A continuous transformation, such as SU(2) or SU(3) ``rotations'' in isospin space, is described by some parameters $\theta^a$ (for SU(2) $a=1,2,3$ whereas for SU(3) $a=1,\ldots,8$). We define these parameters such that, when they vanish, the corresponding transformation is simply the identity transformation (think about a rotation of zero degrees, which means simply ``do no rotation at all''). The element $g(d\theta^a)$ parametrized by infinitesimal ``angles'' $d\theta^a$ corresponds to a transformation infinitesimally close to the identity. If $D(g)$ is a representation of this transformation over some vector space, then $D(g)$ will be infinitesimally close to the identity matrix, and can be Taylor expanded as
\begin{equation*}
    D(g)= \mathds{1} + i d\theta^a \left(-i\dfrac{\partial D(g)}{\partial \theta^a}\biggr|_{\vec{\theta}=0}\right)  + \mathcal{O}(d\theta^2).
    \label{eq:Udt}
\end{equation*}We call 
\begin{equation*}
    T_a=-i\dfrac{\partial D(g)}{\partial\theta^a}\biggr|_{\vec{\theta}=0}
\end{equation*} 
the \textbf{generators} of infinitesimal transformations along the $\theta^a$-direction. 
The factors of $i$ have been introduced so that, for unitary representations, the generators are hermitian matrices, as we saw in sections~\ref{sec:su2inftl} and~\ref{sec:repsSU3} above. Since $\mathds{1}=D(g(\vec{\theta}=0))$, we can write
\begin{equation*}
    T_a = -i\lim_{d\theta^a\to 0} \dfrac{D(g(d\theta^a)) - D(g(0))}{d\theta^a}
\end{equation*}
and we see that $T_a$ can be interpreted as a tangent vector along the $d\theta^a$ direction. This is illustrated in figure~\ref{fig:Lie}.

We argued above that the space of the generators is called the algebra. Now we see that this is also the tangent space at the identity. For this reason we denote it as $T_e G$ in figure~\ref{fig:Lie}: the tangent space to the identity $e$ in group $G$.

It turns out that, for groups of the form of SU($n$), if we are given a generator $T_a$ (i.e. a tangent vector at the identity), we can integrate the entire curve that has $T_a$ as tangent vector at the identity. In other words, for (compact and connected) Lie groups knowing a tangent vector at the identity is enough to know the entire curve\footnote{For general manifolds, i.e. general continuous spaces, to reconstruct a curve we need to know all tangent vectors along the curve, and not just the tangent at one point. Lie groups are special because, apart from being a continuous space, it is also a group.}. To see this, take an element sufficiently close to the identity, which can be written as
\[
    D(g) = \mathds{1} + id\theta\,T
\]
for some generator $T$ in the algebra. Since we know how to compose multiple transformations, cf. eq.~\eqref{eq:homomorphism}, we can keep walking along this curve by multiplying $D(g)$ with itself many times, i.e.
\[
    D(g)^N = \left(\mathds{1} + i d\theta\,T\right)^N.
\]
But this must be equal to some transformation parametrized by $\theta\equiv N\cdot d\theta$ along the $T$ direction,
\[
    D(g(\theta)) = \left(\mathds{1} + i\dfrac{\theta}{N}\,T\right)^N \simeq \exp(i\theta\,T).
\]

This means that any element of the group\footnote{We reinforce that this holds for compact and path-connected groups, such as SU($n$). A non-path-connected group would have elements that cannot be continuously connected to the identity, and of course these points could not be recovered by the exponential map, which is a continuous curve passing through this identity element. Non-compact groups, such as the Lorentz group, also have elements that cannot be written as exponentials. Still, many of the group structures, and especially their representations (which is what interests us in physics) can be recovered by looking at the algebra rather than the whole group.} lying on the curve with tangent $T$ at the identity can be obtained by performing a matrix exponentiation of this generator $T$. It is in this sense that one says that the algebra is related to the group by the $\exp$ map. 

Once we find representations of the algebra (i.e. once we know how states are mapped under infinitesimal transformations), as we did in the main text above, we automatically know how this transformation will take place for any (finite) transformation, since both are related by the $\exp$ map.

%\bibliographystyle{apsrev4-2}
%\bibliography{refs}

\begin{thebibliography}{42}%
\makeatletter
\providecommand \@ifxundefined [1]{%
 \@ifx{#1\undefined}
}%
\providecommand \@ifnum [1]{%
 \ifnum #1\expandafter \@firstoftwo
 \else \expandafter \@secondoftwo
 \fi
}%
\providecommand \@ifx [1]{%
 \ifx #1\expandafter \@firstoftwo
 \else \expandafter \@secondoftwo
 \fi
}%
\providecommand \natexlab [1]{#1}%
\providecommand \enquote  [1]{``#1''}%
\providecommand \bibnamefont  [1]{#1}%
\providecommand \bibfnamefont [1]{#1}%
\providecommand \citenamefont [1]{#1}%
\providecommand \href@noop [0]{\@secondoftwo}%
\providecommand \href [0]{\begingroup \@sanitize@url \@href}%
\providecommand \@href[1]{\@@startlink{#1}\@@href}%
\providecommand \@@href[1]{\endgroup#1\@@endlink}%
\providecommand \@sanitize@url [0]{\catcode `\\12\catcode `\$12\catcode
  `\&12\catcode `\#12\catcode `\^12\catcode `\_12\catcode `\%12\relax}%
\providecommand \@@startlink[1]{}%
\providecommand \@@endlink[0]{}%
\providecommand \url  [0]{\begingroup\@sanitize@url \@url }%
\providecommand \@url [1]{\endgroup\@href {#1}{\urlprefix }}%
\providecommand \urlprefix  [0]{URL }%
\providecommand \Eprint [0]{\href }%
\providecommand \doibase [0]{https://doi.org/}%
\providecommand \selectlanguage [0]{\@gobble}%
\providecommand \bibinfo  [0]{\@secondoftwo}%
\providecommand \bibfield  [0]{\@secondoftwo}%
\providecommand \translation [1]{[#1]}%
\providecommand \BibitemOpen [0]{}%
\providecommand \bibitemStop [0]{}%
\providecommand \bibitemNoStop [0]{.\EOS\space}%
\providecommand \EOS [0]{\spacefactor3000\relax}%
\providecommand \BibitemShut  [1]{\csname bibitem#1\endcsname}%
\let\auto@bib@innerbib\@empty
%</preamble>
\bibitem [{\citenamefont {Halzen}\ and\ \citenamefont
  {Martin}(1984)}]{Halzen:1986}%
  \BibitemOpen
  \bibfield  {author} {\bibinfo {author} {\bibfnamefont {F.}~\bibnamefont
  {Halzen}}\ and\ \bibinfo {author} {\bibfnamefont {A.}~\bibnamefont
  {Martin}},\ }\href@noop {} {\emph {\bibinfo {title} {Quarks and leptons: an
  introductory course in modern particle physics}}}\ (\bibinfo  {publisher}
  {John Wiley \& Sons},\ \bibinfo {year} {1984})\BibitemShut {NoStop}%
\bibitem [{\citenamefont {Griffiths}(2008)}]{Griffiths:2008}%
  \BibitemOpen
  \bibfield  {author} {\bibinfo {author} {\bibfnamefont {D.}~\bibnamefont
  {Griffiths}},\ }\href@noop {} {\emph {\bibinfo {title} {Introduction to
  elementary particles}}},\ \bibinfo {edition} {2nd}\ ed.\ (\bibinfo
  {publisher} {Wiley-VCH Verlag},\ \bibinfo {year} {2008})\BibitemShut
  {NoStop}%
\bibitem [{\citenamefont {Aitchison}\ and\ \citenamefont
  {Hey}(2013)}]{Aitchison:2004cs}%
  \BibitemOpen
  \bibfield  {author} {\bibinfo {author} {\bibfnamefont {I.~J.~R.}\
  \bibnamefont {Aitchison}}\ and\ \bibinfo {author} {\bibfnamefont {A.~J.~G.}\
  \bibnamefont {Hey}},\ }\href@noop {} {\emph {\bibinfo {title} {{Gauge
  Theories in Particle Physics: A Practical Introduction, Volume 2: Non-Abelian
  Gauge Theories : QCD and The Electroweak Theory}}}},\ \bibinfo {edition}
  {4th}\ ed.\ (\bibinfo  {publisher} {Taylor \& Francis},\ \bibinfo {year}
  {2013})\BibitemShut {NoStop}%
\bibitem [{\citenamefont {Gasiorowicz}(1966)}]{Gasiorowicz:1966xra}%
  \BibitemOpen
  \bibfield  {author} {\bibinfo {author} {\bibfnamefont {S.}~\bibnamefont
  {Gasiorowicz}},\ }\href@noop {} {\emph {\bibinfo {title} {{Elementary
  particle physics}}}}\ (\bibinfo  {publisher} {Wiley},\ \bibinfo {address}
  {New York},\ \bibinfo {year} {1966})\BibitemShut {NoStop}%
\bibitem [{\citenamefont {Georgi}(1999)}]{Georgi:1999wka}%
  \BibitemOpen
  \bibfield  {author} {\bibinfo {author} {\bibfnamefont {H.}~\bibnamefont
  {Georgi}},\ }\href@noop {} {\emph {\bibinfo {title} {{Lie algebras in
  particle physics}}}},\ \bibinfo {edition} {2nd}\ ed.,\ Vol.~\bibinfo {volume}
  {54}\ (\bibinfo  {publisher} {Perseus Books},\ \bibinfo {address} {Reading,
  MA},\ \bibinfo {year} {1999})\BibitemShut {NoStop}%
\bibitem [{\citenamefont {Zee}(2016)}]{Zee:2016fuk}%
  \BibitemOpen
  \bibfield  {author} {\bibinfo {author} {\bibfnamefont {A.}~\bibnamefont
  {Zee}},\ }\href@noop {} {\emph {\bibinfo {title} {{Group Theory in a Nutshell
  for Physicists}}}}\ (\bibinfo  {publisher} {Princeton University Press},\
  \bibinfo {address} {USA},\ \bibinfo {year} {2016})\BibitemShut {NoStop}%
\bibitem [{\citenamefont {Thyssen}\ and\ \citenamefont
  {Ceulemans}(2017)}]{Thyssen:2017}%
  \BibitemOpen
  \bibfield  {author} {\bibinfo {author} {\bibfnamefont {P.}~\bibnamefont
  {Thyssen}}\ and\ \bibinfo {author} {\bibfnamefont {A.}~\bibnamefont
  {Ceulemans}},\ }\href@noop {} {\emph {\bibinfo {title} {Shattered Symmetry:
  Group Theory from the Eightfold Way to the Periodic Table}}}\ (\bibinfo
  {publisher} {Oxford University Press},\ \bibinfo {year} {2017})\BibitemShut
  {NoStop}%
\bibitem [{\citenamefont {Heisenberg}(1932{\natexlab{a}})}]{Heisenberg:1932dw}%
  \BibitemOpen
  \bibfield  {author} {\bibinfo {author} {\bibfnamefont {W.}~\bibnamefont
  {Heisenberg}},\ }\href {https://doi.org/10.1007/BF01342433} {\bibfield
  {journal} {\bibinfo  {journal} {Z. Phys.}\ }\textbf {\bibinfo {volume}
  {77}},\ \bibinfo {pages} {1} (\bibinfo {year}
  {1932}{\natexlab{a}})}\BibitemShut {NoStop}%
\bibitem [{\citenamefont {Heisenberg}(1932{\natexlab{b}})}]{Heisenberg:1932II}%
  \BibitemOpen
  \bibfield  {author} {\bibinfo {author} {\bibfnamefont {W.}~\bibnamefont
  {Heisenberg}},\ }\href {https://doi.org/10.1007/BF01337585} {\bibfield
  {journal} {\bibinfo  {journal} {Z. Phys.}\ }\textbf {\bibinfo {volume}
  {78}},\ \bibinfo {pages} {156} (\bibinfo {year}
  {1932}{\natexlab{b}})}\BibitemShut {NoStop}%
\bibitem [{\citenamefont
  {Heisenberg}(1932{\natexlab{c}})}]{Heisenberg:1932III}%
  \BibitemOpen
  \bibfield  {author} {\bibinfo {author} {\bibfnamefont {W.}~\bibnamefont
  {Heisenberg}},\ }\href {https://doi.org/10.1007/BF01335696} {\bibfield
  {journal} {\bibinfo  {journal} {Z. Phys.}\ }\textbf {\bibinfo {volume}
  {80}},\ \bibinfo {pages} {586} (\bibinfo {year}
  {1932}{\natexlab{c}})}\BibitemShut {NoStop}%
\bibitem [{\citenamefont {Kemmer}(1982)}]{Kemmer:1982bj}%
  \BibitemOpen
  \bibfield  {author} {\bibinfo {author} {\bibfnamefont {N.}~\bibnamefont
  {Kemmer}},\ }\href {https://doi.org/10.1051/jphyscol:1982824} {\bibfield
  {journal} {\bibinfo  {journal} {J. Phys. Colloq.}\ }\textbf {\bibinfo
  {volume} {43}},\ \bibinfo {pages} {359} (\bibinfo {year} {1982})}\BibitemShut
  {NoStop}%
\bibitem [{\citenamefont {Cassen}\ and\ \citenamefont
  {Condon}(1936)}]{Cassen:1936dg}%
  \BibitemOpen
  \bibfield  {author} {\bibinfo {author} {\bibfnamefont {B.}~\bibnamefont
  {Cassen}}\ and\ \bibinfo {author} {\bibfnamefont {E.~U.}\ \bibnamefont
  {Condon}},\ }\href {https://doi.org/10.1103/PhysRev.50.846} {\bibfield
  {journal} {\bibinfo  {journal} {Phys. Rev.}\ }\textbf {\bibinfo {volume}
  {50}},\ \bibinfo {pages} {846} (\bibinfo {year} {1936})}\BibitemShut
  {NoStop}%
\bibitem [{\citenamefont {Borrelli}(2017)}]{Borrelli:201781}%
  \BibitemOpen
  \bibfield  {author} {\bibinfo {author} {\bibfnamefont {A.}~\bibnamefont
  {Borrelli}},\ }\href {https://doi.org/10.1016/j.shpsb.2017.03.004} {\bibfield
   {journal} {\bibinfo  {journal} {Studies in History and Philosophy of Science
  Part B: Studies in History and Philosophy of Modern Physics}\ }\textbf
  {\bibinfo {volume} {60}},\ \bibinfo {pages} {81} (\bibinfo {year} {2017})},\
  \bibinfo {note} {on the History of the Quantum, HQ4}\BibitemShut {NoStop}%
\bibitem [{\citenamefont {Feynman}\ \emph {et~al.}(2013)\citenamefont
  {Feynman}, \citenamefont {Leighton},\ and\ \citenamefont
  {Sands}}]{FeynmanLectures}%
  \BibitemOpen
  \bibfield  {author} {\bibinfo {author} {\bibfnamefont {R.}~\bibnamefont
  {Feynman}}, \bibinfo {author} {\bibfnamefont {R.}~\bibnamefont {Leighton}},\
  and\ \bibinfo {author} {\bibfnamefont {M.}~\bibnamefont {Sands}},\
  }\href@noop {} {\emph {\bibinfo {title} {{The Feynman Lectures on Physics
  Vol. III}}}}\ (\bibinfo  {publisher} {Basic Books},\ \bibinfo {address}
  {USA},\ \bibinfo {year} {2013})\BibitemShut {NoStop}%
\bibitem [{\citenamefont {Workman}\ \emph {et~al.}(2022)\citenamefont {Workman}
  \emph {et~al.}}]{ParticleDataGroup:2022pth}%
  \BibitemOpen
  \bibfield  {author} {\bibinfo {author} {\bibfnamefont {R.~L.}\ \bibnamefont
  {Workman}} \emph {et~al.} (\bibinfo {collaboration} {Particle Data Group}),\
  }\href {https://doi.org/10.1093/ptep/ptac097} {\bibfield  {journal} {\bibinfo
   {journal} {PTEP}\ }\textbf {\bibinfo {volume} {2022}},\ \bibinfo {pages}
  {083C01} (\bibinfo {year} {2022})}\BibitemShut {NoStop}%
\bibitem [{\citenamefont {Machleidt}\ and\ \citenamefont
  {Slaus}(2001)}]{Machleidt:2001rw}%
  \BibitemOpen
  \bibfield  {author} {\bibinfo {author} {\bibfnamefont {R.}~\bibnamefont
  {Machleidt}}\ and\ \bibinfo {author} {\bibfnamefont {I.}~\bibnamefont
  {Slaus}},\ }\href {https://doi.org/10.1088/0954-3899/27/5/201} {\bibfield
  {journal} {\bibinfo  {journal} {J. Phys. G}\ }\textbf {\bibinfo {volume}
  {27}},\ \bibinfo {pages} {R69} (\bibinfo {year} {2001})},\ \Eprint
  {https://arxiv.org/abs/nucl-th/0101056} {arXiv:nucl-th/0101056} \BibitemShut
  {NoStop}%
\bibitem [{\citenamefont {Miller}\ \emph {et~al.}(2006)\citenamefont {Miller},
  \citenamefont {Opper},\ and\ \citenamefont {Stephenson}}]{Miller:2006tv}%
  \BibitemOpen
  \bibfield  {author} {\bibinfo {author} {\bibfnamefont {G.~A.}\ \bibnamefont
  {Miller}}, \bibinfo {author} {\bibfnamefont {A.~K.}\ \bibnamefont {Opper}},\
  and\ \bibinfo {author} {\bibfnamefont {E.~J.}\ \bibnamefont {Stephenson}},\
  }\href {https://doi.org/10.1146/annurev.nucl.56.080805.140446} {\bibfield
  {journal} {\bibinfo  {journal} {Ann. Rev. Nucl. Part. Sci.}\ }\textbf
  {\bibinfo {volume} {56}},\ \bibinfo {pages} {253} (\bibinfo {year} {2006})},\
  \Eprint {https://arxiv.org/abs/nucl-ex/0602021} {arXiv:nucl-ex/0602021}
  \BibitemShut {NoStop}%
\bibitem [{\citenamefont {Jenkins}\ \emph {et~al.}(2005)\citenamefont {Jenkins}
  \emph {et~al.}}]{Jenkins:2005jk}%
  \BibitemOpen
  \bibfield  {author} {\bibinfo {author} {\bibfnamefont {D.~G.}\ \bibnamefont
  {Jenkins}} \emph {et~al.},\ }\href
  {https://doi.org/10.1103/PhysRevC.72.031303} {\bibfield  {journal} {\bibinfo
  {journal} {Phys. Rev. C}\ }\textbf {\bibinfo {volume} {72}},\ \bibinfo
  {pages} {031303} (\bibinfo {year} {2005})}\BibitemShut {NoStop}%
\bibitem [{\citenamefont {MacFarlane}\ \emph {et~al.}(1965)\citenamefont
  {MacFarlane}, \citenamefont {Pinski},\ and\ \citenamefont
  {Sudarshan}}]{MacFarlane:1965wp}%
  \BibitemOpen
  \bibfield  {author} {\bibinfo {author} {\bibfnamefont {A.~J.}\ \bibnamefont
  {MacFarlane}}, \bibinfo {author} {\bibfnamefont {G.}~\bibnamefont {Pinski}},\
  and\ \bibinfo {author} {\bibfnamefont {G.}~\bibnamefont {Sudarshan}},\ }\href
  {https://doi.org/10.1103/PhysRev.140.B1045} {\bibfield  {journal} {\bibinfo
  {journal} {Phys. Rev.}\ }\textbf {\bibinfo {volume} {140}},\ \bibinfo {pages}
  {B1045} (\bibinfo {year} {1965})}\BibitemShut {NoStop}%
\bibitem [{\citenamefont {Wohl}(1982)}]{Wohl:1982}%
  \BibitemOpen
  \bibfield  {author} {\bibinfo {author} {\bibfnamefont {C.~G.}\ \bibnamefont
  {Wohl}},\ }\href {https://doi.org/10.1119/1.12743} {\bibfield  {journal}
  {\bibinfo  {journal} {Am. J. Phys.}\ }\textbf {\bibinfo {volume} {50}},\
  \bibinfo {pages} {748} (\bibinfo {year} {1982})}\BibitemShut {NoStop}%
\bibitem [{\citenamefont {Fliagin}\ \emph {et~al.}(1959)\citenamefont
  {Fliagin}, \citenamefont {Dzhelepov}, \citenamefont {Kiselev},\ and\
  \citenamefont {Oganesian}}]{Fliagin:1959}%
  \BibitemOpen
  \bibfield  {author} {\bibinfo {author} {\bibfnamefont {V.~B.}\ \bibnamefont
  {Fliagin}}, \bibinfo {author} {\bibfnamefont {V.~P.}\ \bibnamefont
  {Dzhelepov}}, \bibinfo {author} {\bibfnamefont {V.~S.}\ \bibnamefont
  {Kiselev}},\ and\ \bibinfo {author} {\bibfnamefont {K.}~\bibnamefont
  {Oganesian}},\ }\href@noop {} {\bibfield  {journal} {\bibinfo  {journal}
  {JETP}\ }\textbf {\bibinfo {volume} {8}},\ \bibinfo {pages} {592} (\bibinfo
  {year} {1959})}\BibitemShut {NoStop}%
\bibitem [{\citenamefont {Rochester}\ and\ \citenamefont
  {Butler}(1947)}]{Rochester:1947mi}%
  \BibitemOpen
  \bibfield  {author} {\bibinfo {author} {\bibfnamefont {G.~D.}\ \bibnamefont
  {Rochester}}\ and\ \bibinfo {author} {\bibfnamefont {C.~C.}\ \bibnamefont
  {Butler}},\ }\href {https://doi.org/10.1038/160855a0} {\bibfield  {journal}
  {\bibinfo  {journal} {Nature}\ }\textbf {\bibinfo {volume} {160}},\ \bibinfo
  {pages} {855} (\bibinfo {year} {1947})}\BibitemShut {NoStop}%
\bibitem [{\citenamefont {Brown}\ \emph {et~al.}(1949)\citenamefont {Brown},
  \citenamefont {Camerini}, \citenamefont {Fowler}, \citenamefont {Muirhead},
  \citenamefont {Powell},\ and\ \citenamefont {Ritson}}]{Brown:1949mj}%
  \BibitemOpen
  \bibfield  {author} {\bibinfo {author} {\bibfnamefont {R.}~\bibnamefont
  {Brown}}, \bibinfo {author} {\bibfnamefont {U.}~\bibnamefont {Camerini}},
  \bibinfo {author} {\bibfnamefont {P.~H.}\ \bibnamefont {Fowler}}, \bibinfo
  {author} {\bibfnamefont {H.}~\bibnamefont {Muirhead}}, \bibinfo {author}
  {\bibfnamefont {C.~F.}\ \bibnamefont {Powell}},\ and\ \bibinfo {author}
  {\bibfnamefont {D.~M.}\ \bibnamefont {Ritson}},\ }\href
  {https://doi.org/10.1038/163082a0} {\bibfield  {journal} {\bibinfo  {journal}
  {Nature}\ }\textbf {\bibinfo {volume} {163}},\ \bibinfo {pages} {82}
  (\bibinfo {year} {1949})}\BibitemShut {NoStop}%
\bibitem [{\citenamefont {Hopper}\ and\ \citenamefont
  {Biswas}(1950)}]{Hopper:1950}%
  \BibitemOpen
  \bibfield  {author} {\bibinfo {author} {\bibfnamefont {V.~D.}\ \bibnamefont
  {Hopper}}\ and\ \bibinfo {author} {\bibfnamefont {S.}~\bibnamefont
  {Biswas}},\ }\href {https://doi.org/10.1103/PhysRev.80.1099} {\bibfield
  {journal} {\bibinfo  {journal} {Phys. Rev.}\ }\textbf {\bibinfo {volume}
  {80}},\ \bibinfo {pages} {1099} (\bibinfo {year} {1950})}\BibitemShut
  {NoStop}%
\bibitem [{\citenamefont {Bardon}\ \emph {et~al.}(1958)\citenamefont {Bardon},
  \citenamefont {Lande}, \citenamefont {Lederman},\ and\ \citenamefont
  {Chinowsky}}]{Bardon:1958guy}%
  \BibitemOpen
  \bibfield  {author} {\bibinfo {author} {\bibfnamefont {M.}~\bibnamefont
  {Bardon}}, \bibinfo {author} {\bibfnamefont {K.}~\bibnamefont {Lande}},
  \bibinfo {author} {\bibfnamefont {L.~M.}\ \bibnamefont {Lederman}},\ and\
  \bibinfo {author} {\bibfnamefont {W.}~\bibnamefont {Chinowsky}},\ }\href
  {https://doi.org/10.1016/0003-4916(58)90048-4} {\bibfield  {journal}
  {\bibinfo  {journal} {Annals Phys.}\ }\textbf {\bibinfo {volume} {5}},\
  \bibinfo {pages} {156} (\bibinfo {year} {1958})}\BibitemShut {NoStop}%
\bibitem [{\citenamefont {Okun}(1957)}]{Okun:1957}%
  \BibitemOpen
  \bibfield  {author} {\bibinfo {author} {\bibfnamefont {L.~B.}\ \bibnamefont
  {Okun}},\ }\href@noop {} {\bibfield  {journal} {\bibinfo  {journal} {JETP}\
  }\textbf {\bibinfo {volume} {7}},\ \bibinfo {pages} {322} (\bibinfo {year}
  {1957})}\BibitemShut {NoStop}%
\bibitem [{\citenamefont {Yamaguchi}(1958)}]{Yamaguchi:1958}%
  \BibitemOpen
  \bibfield  {author} {\bibinfo {author} {\bibfnamefont {Y.}~\bibnamefont
  {Yamaguchi}},\ }\href {https://doi.org/10.1143/PTP.19.622} {\bibfield
  {journal} {\bibinfo  {journal} {Prog. Theor. Phys.}\ }\textbf {\bibinfo
  {volume} {19}},\ \bibinfo {pages} {622} (\bibinfo {year} {1958})}\BibitemShut
  {NoStop}%
\bibitem [{\citenamefont {Bastien}\ \emph {et~al.}(1962)\citenamefont
  {Bastien}, \citenamefont {Berge}, \citenamefont {Dahl},\ and\ \citenamefont
  {Ferro-Luzzi}}]{Bastien:1962zz}%
  \BibitemOpen
  \bibfield  {author} {\bibinfo {author} {\bibfnamefont {P.~L.}\ \bibnamefont
  {Bastien}}, \bibinfo {author} {\bibfnamefont {J.~P.}\ \bibnamefont {Berge}},
  \bibinfo {author} {\bibfnamefont {O.~I.}\ \bibnamefont {Dahl}},\ and\
  \bibinfo {author} {\bibfnamefont {M.}~\bibnamefont {Ferro-Luzzi}},\ }\href
  {https://doi.org/10.1103/PhysRevLett.8.114} {\bibfield  {journal} {\bibinfo
  {journal} {Phys. Rev. Lett.}\ }\textbf {\bibinfo {volume} {8}},\ \bibinfo
  {pages} {114} (\bibinfo {year} {1962})}\BibitemShut {NoStop}%
\bibitem [{\citenamefont {Barnes}\ \emph {et~al.}(1964)\citenamefont {Barnes}
  \emph {et~al.}}]{Barnes:1964pd}%
  \BibitemOpen
  \bibfield  {author} {\bibinfo {author} {\bibfnamefont {V.~E.}\ \bibnamefont
  {Barnes}} \emph {et~al.},\ }\href
  {https://doi.org/10.1103/PhysRevLett.12.204} {\bibfield  {journal} {\bibinfo
  {journal} {Phys. Rev. Lett.}\ }\textbf {\bibinfo {volume} {12}},\ \bibinfo
  {pages} {204} (\bibinfo {year} {1964})}\BibitemShut {NoStop}%
\bibitem [{Gell-Mann, M.()}]{GellMann:1962}%
  \BibitemOpen
  Gell-Mann, M.,\ \href@noop {} {\emph {\bibinfo {title} {1962 International
  Conference on High-Energy Physics at CERN}}}\ (\bibinfo {year}
  {1962})\BibitemShut {NoStop}%
\bibitem [{\citenamefont {Glashow}\ and\ \citenamefont
  {Sakurai}(1962)}]{Glashow:1962zza}%
  \BibitemOpen
  \bibfield  {author} {\bibinfo {author} {\bibfnamefont {S.~L.}\ \bibnamefont
  {Glashow}}\ and\ \bibinfo {author} {\bibfnamefont {J.~J.}\ \bibnamefont
  {Sakurai}},\ }\href {https://doi.org/10.1007/BF02771833} {\bibfield
  {journal} {\bibinfo  {journal} {Nuovo Cim.}\ }\textbf {\bibinfo {volume}
  {26}},\ \bibinfo {pages} {622} (\bibinfo {year} {1962})}\BibitemShut
  {NoStop}%
\bibitem [{\citenamefont {Ikeda}\ \emph {et~al.}(1959)\citenamefont {Ikeda},
  \citenamefont {Ogawa},\ and\ \citenamefont {Ohnuki}}]{Ikeda:1959zz}%
  \BibitemOpen
  \bibfield  {author} {\bibinfo {author} {\bibfnamefont {M.}~\bibnamefont
  {Ikeda}}, \bibinfo {author} {\bibfnamefont {S.}~\bibnamefont {Ogawa}},\ and\
  \bibinfo {author} {\bibfnamefont {Y.}~\bibnamefont {Ohnuki}},\ }\href
  {https://doi.org/10.1143/PTP.22.715} {\bibfield  {journal} {\bibinfo
  {journal} {Prog. Theor. Phys.}\ }\textbf {\bibinfo {volume} {22}},\ \bibinfo
  {pages} {715} (\bibinfo {year} {1959})}\BibitemShut {NoStop}%
\bibitem [{\citenamefont {Kalbfleisch}\ \emph {et~al.}(1964)\citenamefont
  {Kalbfleisch} \emph {et~al.}}]{Kalbfleisch:1964zz}%
  \BibitemOpen
  \bibfield  {author} {\bibinfo {author} {\bibfnamefont {G.~R.}\ \bibnamefont
  {Kalbfleisch}} \emph {et~al.},\ }\href
  {https://doi.org/10.1103/PhysRevLett.12.527} {\bibfield  {journal} {\bibinfo
  {journal} {Phys. Rev. Lett.}\ }\textbf {\bibinfo {volume} {12}},\ \bibinfo
  {pages} {527} (\bibinfo {year} {1964})}\BibitemShut {NoStop}%
\bibitem [{\citenamefont {Goldberg}\ \emph {et~al.}(1964)\citenamefont
  {Goldberg} \emph {et~al.}}]{Goldberg:1964zza}%
  \BibitemOpen
  \bibfield  {author} {\bibinfo {author} {\bibfnamefont {M.}~\bibnamefont
  {Goldberg}} \emph {et~al.},\ }\href
  {https://doi.org/10.1103/PhysRevLett.12.546} {\bibfield  {journal} {\bibinfo
  {journal} {Phys. Rev. Lett.}\ }\textbf {\bibinfo {volume} {12}},\ \bibinfo
  {pages} {546} (\bibinfo {year} {1964})}\BibitemShut {NoStop}%
\bibitem [{\citenamefont {Dauber}\ \emph {et~al.}(1964)\citenamefont {Dauber},
  \citenamefont {Slater}, \citenamefont {Smith}, \citenamefont {Stork},\ and\
  \citenamefont {Ticho}}]{Dauber:1964zz}%
  \BibitemOpen
  \bibfield  {author} {\bibinfo {author} {\bibfnamefont {P.~M.}\ \bibnamefont
  {Dauber}}, \bibinfo {author} {\bibfnamefont {W.~E.}\ \bibnamefont {Slater}},
  \bibinfo {author} {\bibfnamefont {L.~T.}\ \bibnamefont {Smith}}, \bibinfo
  {author} {\bibfnamefont {D.~H.}\ \bibnamefont {Stork}},\ and\ \bibinfo
  {author} {\bibfnamefont {H.~K.}\ \bibnamefont {Ticho}},\ }\href
  {https://doi.org/10.1103/PhysRevLett.13.449} {\bibfield  {journal} {\bibinfo
  {journal} {Phys. Rev. Lett.}\ }\textbf {\bibinfo {volume} {13}},\ \bibinfo
  {pages} {449} (\bibinfo {year} {1964})}\BibitemShut {NoStop}%
\bibitem [{\citenamefont {Maki}(1961)}]{Maki:1961}%
  \BibitemOpen
  \bibfield  {author} {\bibinfo {author} {\bibfnamefont {Z.}~\bibnamefont
  {Maki}},\ }\href {https://doi.org/10.1143/PTPS.19.33} {\bibfield  {journal}
  {\bibinfo  {journal} {Prog. Theor. Phys. Suppl.}\ }\textbf {\bibinfo {volume}
  {19}},\ \bibinfo {pages} {33} (\bibinfo {year} {1961})}\BibitemShut {NoStop}%
\bibitem [{\citenamefont {Okun}(2007)}]{Okun:2006nq}%
  \BibitemOpen
  \bibfield  {author} {\bibinfo {author} {\bibfnamefont {L.~B.}\ \bibnamefont
  {Okun}},\ }\href {https://doi.org/10.1143/PTPS.167.163} {\bibfield  {journal}
  {\bibinfo  {journal} {Prog. Theor. Phys. Suppl.}\ }\textbf {\bibinfo {volume}
  {167}},\ \bibinfo {pages} {163} (\bibinfo {year} {2007})},\ \Eprint
  {https://arxiv.org/abs/hep-ph/0611298} {arXiv:hep-ph/0611298} \BibitemShut
  {NoStop}%
\bibitem [{\citenamefont {Gell-Mann}(1961)}]{Gell-Mann:1961TheEightFold}%
  \BibitemOpen
  \bibfield  {author} {\bibinfo {author} {\bibfnamefont {M.}~\bibnamefont
  {Gell-Mann}},\ }\href {https://doi.org/10.2172/4008239} {\emph {\bibinfo
  {title} {The eightfold way: A theory of strong interaction summetry}}}\
  (\bibinfo  {publisher} {California Institute of Technology},\ \bibinfo {year}
  {1961})\BibitemShut {NoStop}%
\bibitem [{\citenamefont {Ne'eman}(1961)}]{Neeman:1961TheEightFold}%
  \BibitemOpen
  \bibfield  {author} {\bibinfo {author} {\bibfnamefont {Y.}~\bibnamefont
  {Ne'eman}},\ }\href
  {https://doi.org/https://doi.org/10.1016/0029-5582(61)90134-1} {\bibfield
  {journal} {\bibinfo  {journal} {Nuclear Physics}\ }\textbf {\bibinfo {volume}
  {26}},\ \bibinfo {pages} {222} (\bibinfo {year} {1961})}\BibitemShut
  {NoStop}%
\bibitem [{\citenamefont {Gell-Mann}(1964)}]{Gell-Mann:1964Quarks}%
  \BibitemOpen
  \bibfield  {author} {\bibinfo {author} {\bibfnamefont {M.}~\bibnamefont
  {Gell-Mann}},\ }\href
  {https://doi.org/https://doi.org/10.1016/S0031-9163(64)92001-3} {\bibfield
  {journal} {\bibinfo  {journal} {Physics Letters}\ }\textbf {\bibinfo {volume}
  {8}},\ \bibinfo {pages} {214} (\bibinfo {year} {1964})}\BibitemShut {NoStop}%
\bibitem [{\citenamefont {Zweig}(1964)}]{Zweig:1964}%
  \BibitemOpen
  \bibfield  {author} {\bibinfo {author} {\bibfnamefont {G.}~\bibnamefont
  {Zweig}}\ }\href {https://doi.org/10.17181/CERN-TH-412}
  {10.17181/CERN-TH-412} (\bibinfo {year} {1964})\BibitemShut {NoStop}%
\bibitem [{\citenamefont {Riordan}(1992)}]{Riordan:1992hr}%
  \BibitemOpen
  \bibfield  {author} {\bibinfo {author} {\bibfnamefont {E.~M.}\ \bibnamefont
  {Riordan}},\ }\href {https://doi.org/10.1126/science.256.5061.1287}
  {\bibfield  {journal} {\bibinfo  {journal} {Science}\ }\textbf {\bibinfo
  {volume} {256}},\ \bibinfo {pages} {1287} (\bibinfo {year}
  {1992})}\BibitemShut {NoStop}%
\end{thebibliography}
%

\end{document}